\documentclass[prx,letterpaper,nobalancelastpage,twocolumn,superscriptaddress,nofootinbib]{revtex4-2}

\usepackage{lineno}
\usepackage{graphicx}
\usepackage{amsmath}
\usepackage{bbold}
\usepackage{amssymb}
\usepackage[english]{babel}
\usepackage{color}
\usepackage[version=4]{mhchem}
\usepackage[hidelinks]{hyperref}
\hypersetup{colorlinks=true,citecolor=blue,linkcolor=blue,urlcolor=blue,pdfstartview=FitH}
\usepackage{dsfont}
\usepackage{placeins}
\usepackage{mathtools}
\usepackage[normalem]{ulem}
\usepackage{lipsum}
\usepackage{comment}
\usepackage{booktabs}
\usepackage{mathtools}
\usepackage{array}
\usepackage{tabularx}

\usepackage{multirow}

\setcitestyle{super}

\usepackage{soul}
\newcommand{\ket}[1]{| #1 \rangle}
\newcommand{\bra}[1]{\langle #1 |}

\newcommand{\ketbra}[1]{|  #1 \rangle \langle #1 |}

\newcommand{\Caltech}{California Institute of Technology, Pasadena, CA 91125, USA}

\usepackage{caption}
\usepackage[caption=false]{subfig}

\DeclareCaptionLabelSeparator{bar}{ \textbf{\textbar}~}
\usepackage{booktabs,tabularx,array}
\newcolumntype{Y}{>{\centering\arraybackslash}X}
\usepackage{enumitem}
\setlist{nolistsep}
\begin{document}

\title{Experimental observation of conformal field theory spectra}

\author{Xiangkai Sun}\thanks{These authors contributed equally to this work}\affiliation{\Caltech}
\author{Yuan Le}\thanks{These authors contributed equally to this work}\affiliation{\Caltech}
\author{Stephen Naus}\thanks{These authors contributed equally to this work}\affiliation{\Caltech}
\author{Richard Bing-Shiun Tsai}\thanks{These authors contributed equally to this work}\affiliation{\Caltech}
\author{Lewis R. B. Picard}\affiliation{\Caltech}
\author{Sara Murciano}\affiliation{Universit\'e Paris-Saclay, CNRS, LPTMS, 91405, Orsay, France.}
\author{Michael Knap}\affiliation{Technical University of Munich, TUM School of Natural Sciences, Physics Department, 85748 Garching, Germany}\affiliation{Munich Center for Quantum Science and Technology (MCQST), Schellingstr. 4, 80799 M{\"u}nchen, Germany}
\author{Jason Alicea}\affiliation{\Caltech}
\author{Manuel Endres}\email{mendres@caltech.edu}\affiliation{\Caltech}

\maketitle

\noindent\textbf{Conformal field theories (CFTs) feature prominently in high-energy physics~\cite{maldacena_large-n_1999,Friedan:1985ge}, statistical mechanics~\cite{BelavinPolyakovZamolodchikov1984,Cardyscalingandrenormalization,yellowbook1997}, and condensed matter~\cite{sachdev_quantum_2011,sachdev_gapless_1993,nayak_non-abelian_2008,kitaev_talk_2015}. For example, CFTs govern emergent universal properties of systems tuned to quantum phase transitions~\cite{sachdev_quantum_2011,holzhey_geometric_1994,calabrese_entanglement_2009,Giamarchi}, including their entanglement, correlations, and low-energy excitation spectra.  
Much of the rich structure predicted by CFTs nevertheless remains unobserved in experiment.  Here we directly observe the energy excitation spectra of emergent CFTs at quantum phase transitions---recovering universal energy ratios characteristic of the underlying field theories~\cite{cardy_conformal_1984,cardy_boundary_1989,Cardy2006BCFT}. Specifically, we develop and implement a modulation technique to resolve a Rydberg chain's finite-size spectra, variably tuned to quantum phase transitions described by either Ising or tricritical Ising CFTs.
We also employ local control to distinguish parities of excitations under reflection and, in the tricritical Ising chain, to induce transitions between distinct CFT spectra associated with changing boundary conditions.
By utilizing a variant of the modulation technique, we furthermore study the dynamical structure factor of the critical system, which is closely related to the correlation of an underlying Ising conformal field.
Our work not only probes the emergence of CFT features in a quantum simulator, but also provides a technique for diagnosing a priori unknown universality classes in future experiments.
}

\vspace{3mm}

\begin{figure*}[t!]
    \centering
    \includegraphics[width=\linewidth]{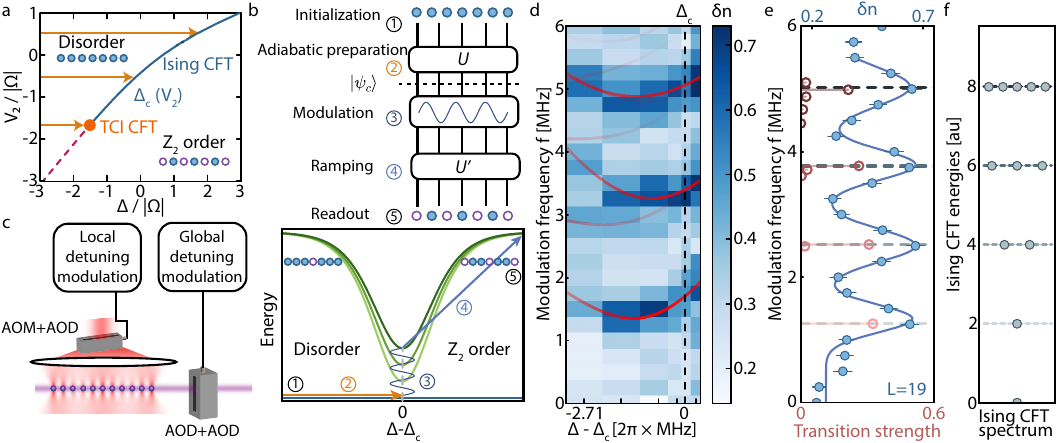}
    \caption{\textbf{Many-body spectroscopy and the Ising CFT spectrum.} 
    \textbf{(a)} Phase diagram of the Fendley-Sengupta-Sachdev Hamiltonian, reproduced from Ref.~\cite{slagle_microscopic_2021}. Arrows represent the adiabatic sweep trajectory, ending at critical points probed in this work. 
    \textbf{(b)} Modulation spectroscopy protocol (top) and corresponding schematic spectral evolution (bottom).  We adiabatically evolve an initial disordered state to Ising criticality and then apply modulation to target a low-lying excited state; the readout of the change of the number of atoms in $\ket{0}$, $\delta n$, is performed following a subsequent detuning sweep to the $\mathbb{Z}_2$-ordered / disordered phase. 
    \textbf{(c)} Modulation spectroscopy implementation setup. We use acousto-optic deflectors (AODs) and an acousto-optic modulator (AOM) to control the global and the local detuning modulation with the Rydberg laser frequency and the tweezer-induced light shift.
    \textbf{(d)} Spectroscopy of the many-body Hamiltonian on a 19-atom array with modulation applied at various detunings $\Delta$. Red solid lines are numerical calculations of the excited state energies of Eq.~\eqref{eq:masterH}, with the opacity indicating the transition strength between the ground state and each excited state. The dashed line represents the numerically determined critical detuning $\Delta_c$. 
    \textbf{(e)} Spectroscopy at the critical detuning $\Delta_c$ of a 19-atom chain. We observe an increase in the number of excitations after the sweep ($\delta n$) (blue, top axis) at frequencies coinciding with excited state energies. The solid line is a fit with a sum of four Gaussians and indicate fit range. Dashed lines indicate the CFT predicted energy ratio. Red circles are numerical calculations of transition strength $|\bra{g}\hat{K}\ket{e}|^2$ (bottom axis). The color code represents clusters of states with close eigenenergies which will become degenerate in the thermodynamic limit as predicted by the Ising CFT.
    \textbf{(f)} The predicted even-parity Ising CFT spectrum, from exact analytical calculations. The Ising CFT energy levels have the universal ratio 2:4:6:8. }
    \label{fig:isingspectroscopy}
\end{figure*}

Conformal field theories (CFTs) are quantum field theories preserving conformal transformations that tightly constrain physical observables. 
Although conformal symmetry generally does not manifest microscopically---e.g., it is exact in neither the Standard Model nor in typical many-body Hamiltonians---it can emerge over appropriate length scales. In the high-energy context, the AdS/CFT correspondence posits a duality between a CFT and quantum gravity in anti–de Sitter space~\cite{maldacena_large-n_1999}; the error-correcting code of Pastawski, Yoshida, Harlow, and Preskill provides a concrete microscopic setting that reproduces key features of this correspondence~\cite{pastawski_holographic_2015}. In many-body systems, CFTs commonly describe universal phenomena at quantum phase transitions (QPTs)---zero-temperature critical points separating distinct phases of matter, at which conformal symmetry emerges at long distances and low energies~\cite{sachdev_quantum_2011}.   

A CFT is fixed by a set of data from which one can infer universal features such as the spatial structure of quantum entanglement and critical exponents governing correlations in space-time.
These universal features manifest themselves in finite-size scaling, which become a powerful diagnostic tool in systems governed by two-dimensional CFTs: 
A strategy outlined by Cardy in the 1980s~\cite{cardy_conformal_1984,Cardy1986} showed that the pattern of low-energy levels in a finite system directly reveals the operator content of the CFT---that is, which fields are present along with their scaling dimensions.  Moreover, toggling boundary conditions acts as a control knob that filters the allowed operators and hence alters the observable low-energy spectrum.

In quantum materials, candidate platforms for probing CFTs include magnetic insulators and quantum Hall systems, which admit probes such as inelastic neutron scattering~\cite{tennant_unbound_1993,werner2001lineshapes,lake_quantum_2005,hirtenlechner2014neutron,povarov_scaling_2015,lake_multispinon_2013,mourigal_fractional_2013} and electrical and thermal transport~\cite{grayson_continuum_1998,chang_observation_1996,banerjee_observation_2018,cohen_universal_2023} capable of revealing select CFT properties. 
Recent advances in programmable analog and digital quantum simulators have enabled probing QPTs~\cite{greiner_quantum_2002,bakr_probing_2010,endres_higgs_2012,bernien_probing_2017,zhang_observation_2017} with complementary manipulation and readout techniques that offer a pathway to refined experimental characterization of CFTs.  In this context, experiments have attempted the ground-state preparation of quantum critical periodic Rydberg chains~\cite{fang_probing_2024}. Furthermore, dynamical properties---e.g., Kibble--Zurek scaling and late-time quantum coarsening---were probed when ramping through QPTs~\cite{keesling_quantum_2019,ebadi_quantum_2021,manovitz_quantum_2025,andersen_thermalization_2025}. 
However, a direct measurement of the rich excitation spectra of CFTs is outstanding, in either solid-state systems or quantum simulators, despite its fundamental relevance.

Here we use a programmable neutral atom quantum simulator to measure the finite-size many-body excitation spectrum at one-dimensional QPTs, which are governed by two-dimensional (1+1D space-time) Ising and tricritical Ising (TCI) CFTs.
To this end, we develop a modulation spectroscopy technique that implements coherent control between many-body states residing in targeted symmetry sectors. In the Ising case, we experimentally resolve excitation energies whose ratios and scaling with system size agree well with universal predictions of the Ising CFT in open chains.  
At the TCI point, we further probe the dependence of the spectrum on boundary conditions by applying additional local detunings at the chain's edges---revealing signatures consistent with the TCI CFT subject to three different fixed-point boundary conditions.  
To overcome the difficulty of identifying individual excitation energies in large systems due to the large density of states, we utilize a variant of the modulation spectroscopy to measure the dynamical structure factor of a system tuned to the Ising criticality, which serves as an experimental approach to probe the universal scaling function of the underlying field correlation~\cite{sachdev_quantum_2011}. 
This work establishes modulation spectroscopy in quantum simulators as a powerful tool to study general QPTs, including in regimes where classical simulations are challenging and the critical properties may be unknown a priori.

Our experiments are performed with a one-dimensional lattice of $L$ interacting Rydberg atoms~\cite{madjarov_high-fidelity_2020}.  
Each qubit is encoded between a ground state $\ket{0}$ and a Rydberg state $\ket{1}$. These states are coupled with a global Rabi frequency $\Omega$ and detuning $\Delta$. When a pair of atoms at positions $r_i,r_j$ are both in $\ket{1}$, they interact via a native van der Waals potential $V_{ij} \propto |r_i-r_j|^{-6}$. We work in the Rydberg blockade regime in which $V_1\equiv V_{i,i+1} \gg \Omega, \Delta$, effectively forbidding neighboring atoms from simultaneously entering the $\ket{1}$ state. Additionally, we can apply site-dependent detunings $\delta \Delta_i$~\cite{chen_continuous_2023,de_oliveira_demonstration_2025,manovitz_quantum_2025} (Methods). The effective Hamiltonian we use to describe our setup is
\begin{equation}\label{eq:masterH}
\begin{split}
    \hat{H} = &\sum_i \left[\dfrac{\Omega}{2}\hat{P}_{i-1}\hat{X}_i\hat{P}_{i+1} - (\Delta+\delta\Delta_i) \hat{n}_i\right] \\
    + &\sum_{j-i\ge 2}V_{ij}\hat{n}_i \hat{n}_j + \hat{H}_2,
\end{split}
\end{equation}
where $\hat{P}_i=\ket{0}_i\bra{0}_i$, $\hat{X}_i=\ket{1}_i\bra{0}_i+\ket{0}_i\bra{1}_i$, $\hat{n}_i=\ket{1}_i\bra{1}_i$ with $i=1,2,...,L$, and $\hat{H}_2$ contains higher-order corrections due to finite $V_1$ (Methods). Hereafter we take $\delta \Delta_i = 0$ until specified otherwise.

The above Hamiltonian reduces to the Fendley-Sengupta-Sachdev~\cite{fendley_competing_2004} (FSS) model upon taking $V_1 \rightarrow \infty$, retaining finite next-nearest-neighbor interaction $V_2 \equiv V_{i,i+2}$, and neglecting further-neighbor interactions. In this simplified limit, the corresponding phase diagram features a trivial disordered phase and a $\mathbb{Z}_2$-ordered charge density wave (CDW) phase (Fig.~\ref{fig:isingspectroscopy}a). These phases are separated by an extended second-order transition line, described by the Ising CFT, that at some $V_2<0$ terminates at a tricritical point, described by the TCI CFT, before becoming first-order (dashed line in Fig.~\ref{fig:isingspectroscopy}a). Using quasi-adiabatic sweeps \cite{bernien_probing_2017}, we prepare our system close to the ground state in the vicinity of the phase transition. The native van der Waals interaction gives access to the repulsive regime with $V_2>0$. We reach effective attractive $V_2<0$, necessary for probing the TCI point, with an inverted sweep protocol, targeting the most excited state of the system in the Rydberg-blockaded regime instead of the ground state (Methods).

\vspace{3mm}
\noindent\textbf{Many-body modulation spectroscopy}

To probe the excitation spectrum close to the critical point, we develop a modulation spectroscopy technique (the \textit{modulation-ramp-probe sequence}). In short, we prepare the ground state close to a critical detuning, $\Delta_c$, apply a modulation to couple to excited states, ramp into either the $\mathbb{Z}_2$ or disordered phase, and then probe the population transferred into excited states (Fig.~\ref{fig:isingspectroscopy}b). 

We perform two types of modulations, modulating the global detuning $\Delta$ and local detunings $\delta \Delta_i$ (Fig.~\ref{fig:isingspectroscopy}c, Methods). 
Global detuning modulation perturbs the Hamiltonian via
\begin{equation}
\delta \hat{H}_{\rm global}(t) = A \cos(2\pi f t+\varphi) \ \mathrm{e}^{-\frac{t^2}{w^2}}\sum_i \hat{n}_i,
\end{equation}
where $A$ is the modulation amplitude, $f$ is the modulation frequency, $\varphi$ is the initial phase of the modulaion, and $w$ is the width of a Gaussian envelope. 

We read out the population transfer by measuring the number of atoms in $\ket{0}$ in the final state, $\hat{n}_g= \sum_i \ket{0}_i\bra{0}_i$, after ramping to either the $\mathbb{Z}_2$ ordered phase or the disordered phase. Transition to low-energy excited states appears as a change in $\langle \hat{n}_g\rangle$ relative to that of the many-body ground state, which we denote as $\delta n$. For a small system with $L =7$, 
we observe temporal oscillation of $\delta n$, indicating coherent oscillation between the ground state and a chosen excited state (Methods, Ext. Data Fig.~\ref{fig:RabiRamsey}). 

To probe the spectrum of larger systems, we drive the system with a weak drive for less than half a Rabi oscillation cycle. In this regime, $\delta n$ is given approximately as a sum of Gaussians located at energy differences between ground and low-energy excited states and weighted with transition matrix elements (Methods):
\begin{equation}\label{eq:peakHeight}
    \delta n\propto A^2w^2 \sum_{e\neq g} e^{-2\pi^2 w^2 \left(\frac{E_e-E_g}{h} - f\right)^2} \left|\bra{g}\hat{K}\ket{e}\right|^2,
\end{equation}
where $h$ is Planck's constant, $\hat{K}\equiv \sum_i \hat{n}_i$ is the modulation operator, $\ket{g}$ is the many-body ground state with energy $E_g$, and $\ket{e}$ runs over many-body excited states with energy $E_e$.  

Equation~\eqref{eq:peakHeight} reveals that when $f$ coincides with the energy difference between the ground state and an excited state, population is transferred with a rate proportional to the square of the transition matrix element---which, importantly, is constrained by reflection symmetry.  In particular, under spatial reflection all Hamiltonian eigenstates either remain invariant (even parity) or acquire a $(-1)$ factor (odd parity).  Since both the ground state and the global modulation have even parity, the transition matrix element to odd-parity excited states vanishes.  Local detuning modulation detailed below allows us to additionally access the odd-parity sector.  

\begin{figure}[t!]
    \centering
    \includegraphics[width=1\linewidth]
    {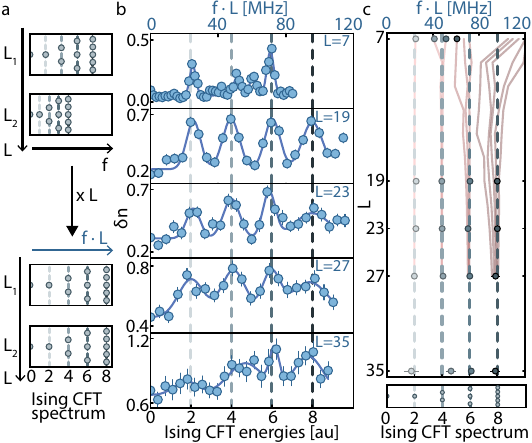}
    \caption{\textbf{Scaling of the finite-size spectrum.} 
    \textbf{(a)} Sketch of scaling modulation frequency with system size. When multiplying the frequency with the system size, the eigenvalues for different system sizes collapse and yield a universal ratio predicted by the CFT.
    \textbf{(b)} Spectroscopy at the critical detuning $\Delta_c$ at various system sizes. In the horizontal axis the modulation frequency $f$ is scaled by system size $L$. Dashed lines are numerically extrapolated eigenvalues, corresponding to the even-parity Ising CFT spectrum with the universal ratio 2:4:6:8. Solid lines are multi-Gaussian fits to the data and indicate fit ranges.  
    \textbf{(c)} First four spectral peak positions $f$, rescaled by $L$, for each system size. Solid lines are the numerically calculated rescaled even-parity state energies with the same colors as in Fig.~\ref{fig:isingspectroscopy}e. Dashed lines are the extrapolated rescaled energies. }
    \label{fig:isingscaling}
\end{figure}

\vspace{3mm}
\noindent\textbf{Observation of the Ising CFT spectrum}

Using this global modulation technique, we first probe the even-parity excitation spectrum close to the critical point for positive interaction strength $V_2/\Omega = 0.51$, whose critical behaviors are described by the Ising CFT, in a chain of length $L=19$ (Fig.~\ref{fig:isingspectroscopy}d, \ref{fig:isingspectroscopy}e). We quasi-adiabatically prepare the system close to its ground state at variable $\Delta$, including the critical detuning $\Delta_c$ corresponding to the numerically predicted critical point in the thermodynamic limit (Methods, Ext. Data Fig.~\ref{fig:isingScalingCollapse}).
For each $\Delta$, we scan the global  modulation frequency $f$. We find increased population transfer $\delta n$ (blue shading in Fig.~\ref{fig:isingspectroscopy}d) at modulation frequencies that coincide with numerically calculated excited-state energies (red lines in Fig.~\ref{fig:isingspectroscopy}d). In addition, the population transfer correlates with the numerically determined transition strength $| \bra{g}\hat{K}\ket{e}|^2$ (opacity of red lines in Fig.~\ref{fig:isingspectroscopy}d). 
These results show that we can prepare the system close to the ground state and can resolve individual excited-state energies, in line with exact numerical predictions. 

To study the CFT spectrum, hereafter we focus on the excitation energies measured at the critical point, $\Delta = \Delta_c$. 
We spectrally resolve four peaks, the center frequencies of which are consistent with four eigenstate energies exhibiting the strongest transition strengths (Fig.~\ref{fig:isingspectroscopy}e). 
These four peaks correspond to four energy levels of the even-parity Ising CFT spectrum with universal energy ratio 2:4:6:8~\cite{cardy_boundary_1989,iino_boundary_2020} (Fig.~\ref{fig:isingspectroscopy}f, Methods, Table~\ref{tab:parity-side-by-side}). Despite the degeneracy in the Ising CFT spectrum, it is lifted in a finite-size system by irrelevant perturbations, and only one of the ideally degenerate states strongly couples to the ground state under global modulation (Fig.~\ref{fig:isingspectroscopy}e).

The CFT spectrum energies are predicted to scale inversely with the system size $L$; hence, upon multiplying the excitation frequencies by $L$, peak positions for different system sizes should line up (Fig.~\ref{fig:isingscaling}a). We probe this scaling behavior experimentally by repeating the global modulation spectroscopy at the critical detuning $\Delta_c$ for various chain lengths, ranging from $L=7$ to $35$, and comparing the resulting spectra viewed as a function of $f \cdot L$ (Fig.~\ref{fig:isingscaling}b). 
Given Eq.~\eqref{eq:peakHeight}, we fit these spectra to a sum of Gaussian functions and extract the peak positions---for $L \geq 19$ again obtaining consistency with the universal 2:4:6:8 Ising CFT ratio for the first four even-parity levels (Fig.~\ref{fig:isingscaling}c).
The $L = 7$ spectrum does not, however, collapse to the universal ratio, because even the second even-parity excited state energy exceeds the Rabi frequency and therefore lies outside of the validity regime of the low-energy CFT description.

\begin{figure}[t!]
    \centering
    \includegraphics[width=1\linewidth]{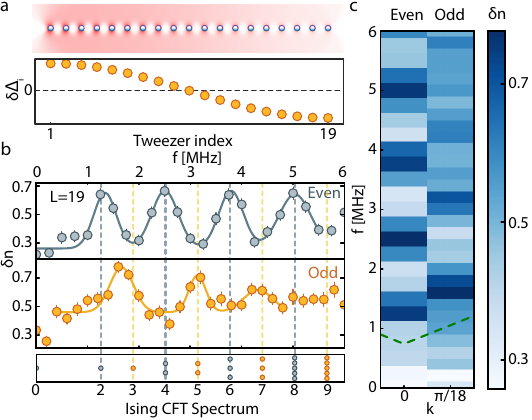}
    \caption{\textbf{Parity-resolved Ising CFT spectrum with local control.} 
    \textbf{(a)} The applied odd-parity local detuning modulation profile with a wavevector of $k=\pi/18$.
    \textbf{(b)} Even- and odd-parity spectra of a 19-atom array tuned to Ising criticality. The measured spectra are consistent with the universal energy ratio, 2:3:4:5:6:7:8, predicted by the Ising CFT. Solid lines are Gaussian fits to the data and indicate fit ranges. Dashed lines are rescaled Ising CFT predictions. 
    \textbf{(c)} Momentum-resolved spectrum. The previous measurements can be interpreted as the even-sector at $k=0$ and the odd-sector at $k=\pi/18$. The slope of the dashed line represents the numerically determined light-cone velocity.
    }
    \label{fig:isingLocal}
\end{figure}

To additionally probe the odd-parity spectrum, we combine global and local detuning modulations to generate an overall modulation pattern that is antisymmetric under reflection~(Fig.~\ref{fig:isingLocal}a).  
Specifically, we realize (Methods)
\begin{equation}
    \delta \hat{H}(t) = A [1+\cos(2\pi f t + \varphi)] e^{-\frac{t^2}{w^2}}\sum_i \cos\left[\dfrac{\pi (i-1)}{L-1}\right] \hat{n}_i.
\end{equation}
Under this odd-parity modulation, a distinct set of spectral peaks emerges, positioned between the even-parity peaks (Fig.~\ref{fig:isingLocal}b). These peaks are consistent with the odd-parity Ising CFT spectrum that exhibits the universal energy ratio of 3:5:7 (Methods, Table~\ref{tab:parity-side-by-side}). 

The universal energy ratios observed experimentally arise from massless, linearly dispersing Majorana fermion excitations that emerge at the Ising phase transition (Methods).  In particular, the Ising CFT predicts emergent fermions carrying excitation energy $E_n = \hbar v k_n$ with $n\in \mathbb{Z}$; here $v$ is a non-universal velocity and $k_n = \frac{\pi}{L}(n+1/2)$ is a finite-size-quantized momentum for an open length-$L$ chain.  Physical states arise from applying an even number of Majorana fermion operators to the ground state---yielding the levels, degeneracy, and reflection properties shown for $L = 19$ at the bottom of Fig.~\ref{fig:isingLocal}b.  A closely related pattern of levels emerges for even-$L$ chains, which we also detect experimentally (Methods, Ext. Data Fig.~\ref{fig:EvenN}).  The linear dispersion, together with fermion filling constraints, are crucial for the universal ratios that we resolve, and thus our experiment constitutes strong evidence for both key features of the Ising CFT.  Further evidence for linear dispersion of emergent fermions arises from modulating with a finite wavevector $k$, where the system only has a strong response when the modulation frequency $f$ exceeds a threshold $f_{th}$ that is linear in $k$ (Methods, Ext.~Data Fig.~\ref{fig:EnergyDispersion}).
From measurements at different $k$, one can extract the non-universal light-cone velocity $v=2\pi |\mathrm{d}f_{th}/\mathrm{d}k|$. Our parity-resolved spectrum in Fig.~\ref{fig:isingLocal}b can be interpreted as observing the onset of the linear dispersion that agrees well with the numerically calculated velocity (Fig.~\ref{fig:isingLocal}c).

Next we deploy similar methods to probe an inherently strongly interacting CFT that defies a free-particle description.

\vspace{3mm}
\noindent\textbf{Approach to the tricritical point}

\begin{figure}[t!]
    \centering
    \includegraphics[width=1\linewidth]{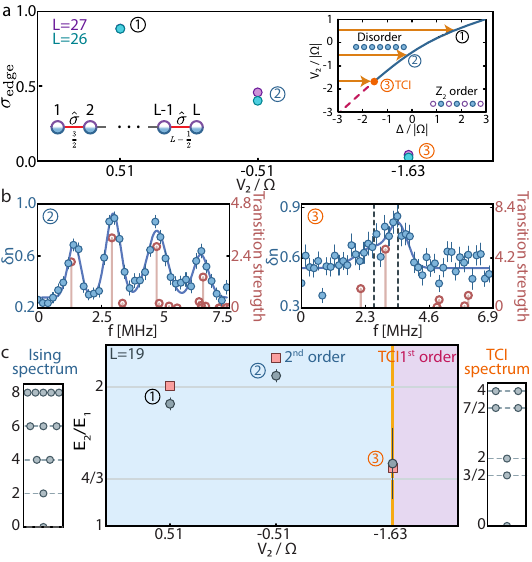}
    \caption{\textbf{Approach to the TCI point.} 
    \textbf{(a)} Boundary field $\sigma_{\rm edge}$ versus interaction strength. The boundary $\sigma_{\rm edge}$ diminishes towards the TCI point, showing the transition from the Ising CFT to the TCI CFT. The right inset (same as Fig.~\ref{fig:isingspectroscopy}a) shows three phase transition configurations and their positions on the phase boundary, and the left inset is a sketch of the site- and bond-labeling of the chain.
    \textbf{(b)} Even-parity spectra at $V_2/\Omega<0$ of a 19-atom array. We fit the spectra (blue, left axis) with a sum of Gaussians (solid lines). For the tricritical spectrum, centers of Gaussians are indicated by dashed lines. Red circles are numerical calculations of transition strength $|\bra{g}\hat{K}\ket{e}|^2$ (right axis).
    \textbf{(c)} Energy ratio $E_{2}/E_{1}$ between the second and the first even-parity excited states. Grey circles are extracted from experimental spectroscopy data. Red squares are extracted from numerical calculations of the first two strong transitions. The Ising CFT spectrum with `fixed' boundary condition (left) and the TCI spectrum with `free' boundary condition (right) yield $E_{2}/E_{1} = 2$ and $4/3$, respectively. 
    }
    \label{fig:BeyondIsing}
\end{figure}

The TCI CFT, which emerges at the termination of the second-order phase boundary in Fig.~\ref{fig:isingspectroscopy}a, carries a distinct set of fields and scaling dimensions compared to the Ising CFT.  
Accordingly, the finite-size energy spectrum at the TCI point admits different universal ratios that, in turn, deeply relate to boundary conditions.  
In our setting, the Ising CFT exhibits a single stable boundary condition---conventionally denoted by `fixed'---at which the edges strongly catalyze CDW order that bleeds into the bulk.  The TCI CFT, in contrast, not only admits stable fixed boundary conditions, but also stable `free' boundary conditions wherein CDW order nucleated by the edges is sharply suppressed.  One can diagnose which boundary condition appears by measuring the CDW order parameter and, more directly, by spectroscopically resolving their characteristic universal energy ratios.    

We experimentally probe the evolution from the Ising to TCI CFT by measuring chains at critical detunings with attractive
next-nearest-neighbor interactions $V_2/\Omega = -0.51$ and $-1.63$ (configurations 2 and 3 in Fig.~\ref{fig:BeyondIsing}a inset). 
The latter configuration is predicted to realize the TCI point based on numerical simulations that determine the $V_2$ at which the first few excited state energy ratios agree with TCI CFT predictions (Methods). 
First, we assess  
CDW order nucleated at the edges of $L = 26, 27$ chains directly after quasi-adiabatic state preparation, i.e., without modulating.
The operator 
$\hat{\sigma}_{i+1/2} = (-1)^{i+1} (\hat{n}_i - \hat{n}_{i+1})$ defines a local CDW order parameter, which we measure at the boundaries to obtain  $\sigma_\text{edge} \equiv \left(|\langle\hat{\sigma}_{3/2}\rangle| + |\langle\hat{\sigma}_{L-1/2}\rangle| \right)/2$ at configurations 1, 2, and 3 in Fig.~\ref{fig:BeyondIsing}a. Upon tiptoeing towards the TCI point, we observe a reduction and eventual nearly complete collapse of edge CDW order---consistent with the emergence of a TCI CFT with free boundary conditions (Fig.~\ref{fig:BeyondIsing}a main panel).  

For further evidence, we perform global modulation spectroscopy on an $L = 19$ chain. With attractive $V_2/\Omega =  -0.51$, the even-parity spectrum remains consistent with the Ising CFT prediction (Fig.~\ref{fig:BeyondIsing}b, left). As a metric, the experimentally extracted ratio between the second and first even-parity excited-state energies is $E_2/E_1 = 2.08(4)$, in agreement with the expected universal Ising CFT ratio of 2. At the TCI point, by contrast, we observe only one skewed spectral feature, which can be fit with a sum of two Gaussians whose centers we associate with even-parity energies $E_1$ and $E_2$ (Fig.~\ref{fig:BeyondIsing}b, right).  Here the extracted ratio is $E_2/E_1 = 1.45(25)$---in line with our simulations (red data points in Fig.~\ref{fig:BeyondIsing}b,c) as well as the free-boundary-condition TCI CFT that predicts $E_2/E_1=4/3$ (Fig.~\ref{fig:BeyondIsing}c, Methods, Table~\ref{tab:TCI_table}).  

Emergence of a TCI CFT with free boundary conditions indicates that explicit translation symmetry breaking from the edges by itself is ineffective at injecting CDW order at the TCI point.  As we show next, we can nevertheless experimentally control the TCI CFT boundary conditions by modifying the Hamiltonian in a way that further promotes CDW order at the chain's edges.

\vspace{3mm}
\noindent\textbf{Boundary tuning of TCI CFT}
\begin{figure}[t!]
    \centering
    \includegraphics[width=1\linewidth]{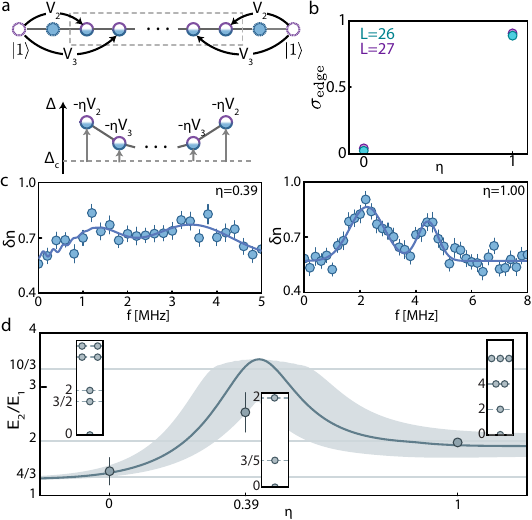}
    \caption{\textbf{Probing different fixed-point boundary conditions for TCI.}
    \textbf{(a)} Controlling the boundary condition with local detunings. We apply local detunings to each atom to mimic the effect of ancillary spins on the boundary. The overall detuning strength is rescaled by a dimensionless parameter $\eta$.
    \textbf{(b)} Boundary field $\sigma_{\rm edge}$ versus the boundary detuning strength $\eta$. A nontrivial $\sigma_{\rm edge}$ emerges in the presence of boundary detuning $\eta=1$, suggesting that the boundary detuning has generated the fixed boundary condition. 
    \textbf{(c)} Spectra at $\eta = 0.39$ and $\eta = 1.00$ of an $L=19$ array. Solid lines are a fit to the theory model (Methods) used to extract excited-state energies.
    \textbf{(d)} The excited state energy ratio of an $L=19$ array under different boundary detunings. Insets are corresponding spectra of the TCI CFT with three different boundary conditions (free, intermediate, and fixed). These three boundary conditions give $4/3, 10/3, 2$, respectively. We experimentally extract the ratio from the spectrum shown in (c). The solid line is the result of the numerical calculation, with the shaded area indicating the uncertainty due to $V_2$. 
    }
    \label{fig:TCI}
\end{figure}

Here we demonstrate that controlling the boundary detunings on the Rydberg array reveals additional universal features of the TCI CFT.  We can ensure maximal edge CDW order by appending the chain with two ancillary atoms on each end \cite{slagle_microscopic_2021}---pinning the outermost ancillas to $\ket{1}$ and, because of Rydberg blockade, the adjacent ancillas to $\ket 0$ (Fig.~\ref{fig:TCI}a, top).  Although these ancillary atoms are completely frozen, the pinned $\ket{1}$ sites impact the `active' degrees of freedom in the remainder of the chain through van der Waals interactions.  These interactions can be exactly emulated in a chain \emph{without} the ancillary CDW-seeding atoms by applying a spatially varying detuning perturbation
\begin{equation}\label{eq:boundaryDetuning}
    \delta \Delta_i = -64V_2 \left[(i+1)^{-6} + (L+2-i)^{-6}\right]
\end{equation}
for site $i$ (Fig.~\ref{fig:TCI}a, bottom, and Ext. Data Fig.~\ref{fig:localdetuningcalib}). 

We are interested in experimentally assessing the fate of the free-boundary-condition TCI CFT upon smoothly turning on $\delta \Delta_i$.  
Thus we prepare the system close to 
the ground state of the Hamiltonian
\begin{equation}
    \hat H_\eta = \hat H_\text{TCI}- \eta \sum_i\delta \Delta_i\ \hat{n}_i,
    \label{eq:boundarydetuningH}
\end{equation}
where $\hat H_\text{TCI}$ corresponds to Eq.~\eqref{eq:masterH} evaluated at the TCI point and $\eta \in [0,1]$ parametrizes the strength of the boundary detuning profile.  First, we consider the limit $\eta = 1$, which corresponds to maximally ``on'' boundary detuning.  Measurements of $\sigma_{\rm edge}$ for $L = 26$ and $27$ chains yield a value close to 1 (Fig.~\ref{fig:TCI}b), strongly suggesting that our boundary detuning has generated fixed boundary conditions.  Modulation spectroscopy targeting even-parity excited states on an $L = 19$ array further supports this assertion: The extracted $E_2/E_1 = 1.98(6)$ value is consistent with ratio of 2 predicted by the fixed-boundary-condition TCI CFT (Fig.~\ref{fig:TCI}c right,d).

The stable free and fixed TCI boundary conditions, respectively demonstrated experimentally at $\eta = 0$ and $\eta = 1$, are separated by an unstable fixed point boundary condition that admits yet another characteristic universal energy spectrum.  
We locate the unstable fixed point by numerically simulating an $L = 19$ array and tracking the energy ratio $E_2/E_1$ of the first two even-parity states as a function of $\eta$ (solid line in Fig.~\ref{fig:TCI}d). At $\eta \equiv \eta_c \approx 0.39$, we find best agreement with the analytical prediction $E_2/E_1 = 10/3$ for the unstable TCI CFT boundary condition. We then experimentally measure the spectrum at $\eta_c$ (Fig.~\ref{fig:TCI}c left) and extract the energy ratio $E_2/E_1 = 2.5(4)$---significantly exceeding the value for both the free and the fixed boundary conditions (Fig.~\ref{fig:TCI}d). The discrepancy between the simulated and measured ratios is primarily attributed to the experimental uncertainty of $V_2$ due to the close atomic distance. We include the uncertainty of $V_2$ ($1.7\%$) in the numerical simulation and find that our measurement falls in the region accounting for the uncertainty (Fig.~\ref{fig:TCI}d, shaded area). 

The TCI CFT also predicts universal ratios among the energy gaps $E_1$ obtained with free, intermediate, and fixed boundary conditions: $E_1^{\rm intermediate}/E_1^{\rm free} = 2/5$ and $E_1^{\rm fixed}/E_1^{\rm free} = 4/3$ (Methods, Table.~\ref{tab:TCI_table}).  Our spectroscopically measured gaps yield $E_1^{\rm intermediate}/E_1^{\rm free} = 0.57(15)$ (close to the CFT prediction) and $E_1^{\rm fixed}/E_1^{\rm free} = 0.91(15)$ (farther, but within the range accounting for $V_2$ uncertainty) (Ext. Data Fig.~\ref{fig:quench}). Complementary gap measurements following a quench to the TCI point further support the gap trends seen in spectroscopy (Methods, Ext. Data Fig.~\ref{fig:quench}).   

At the Ising phase transition line with fixed boundary conditions, we accessed only bosonic excitations arising from the action of an even number of emergent fermion operators on the ground state.  While the fixed-boundary-condition spectrum for the TCI CFT similarly arises from even numbers of (interacting) fermions, the free and intermediate boundary conditions reveal richer field content associated with both bosonic \emph{and} non-bosonic operators. Indeed the fractional $E_2/E_1$ and energy gap ratios backed out experimentally descend from scaling dimensions for non-bosonic TCI fields (Methods)---exemplifying the nontrivial CFT data that our modulation spectroscopy approach can extract.  

\vspace{3mm}
\noindent\textbf{Universal dynamical response in large systems}

\begin{figure*}[htbp]
    \centering
    \includegraphics[width=1\linewidth]{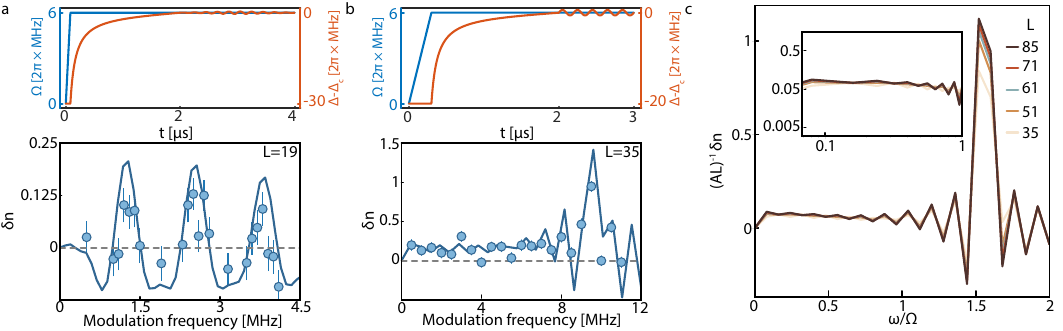}
    \caption{\textbf{Linear response and universal scaling of the dynamical structure factor.}
    \textbf{(a)} Peak-resolved response $\delta n$ under a Gaussian-envelope modulation on an $L=19$ array. (Top) Experimental sweep. The experiment consists of an adiabatic sweep from the disordered phase to the critical point, followed by a Gaussian envelop modulation. (Bottom) The measured response. The solid line is a tensor network simulation of the response of the Fendley-Sengupta-Sachdev model, with peak energies rescaled to match the measured eigenenergies of the Rydberg Hamiltonian.
    \textbf{(b)} Response in the continuum limit under a square-pulse modulation on an $L=35$ array. (Top) Experimental sweep. The modulation applied at the critical point has a square pulse envelop. (Bottom) The measured response. We observe a finite, almost constant response of $\delta n$ at low modulation frequencies. The solid line represents the numerical simulation of the Fendley-Sengupta-Sachdev model using tensor networks.
    \textbf{(c)} Scaling collapse of the response under a square pulse modulation. By rescaling the response with $(AL)^{-1}$, low-energy responses of different system sizes collapse and are nearly constant at low frequencies (inset).
    }
    \label{fig:dynamicSusceptibility}
\end{figure*}

Although we have successfully employed the modulation-ramp-probe sequence to extract universal finite-size CFT spectra, resolving individual levels becomes harder in larger systems: energy spacings between excited states decrease as $L^{-1}$, so the modulation spectroscopy would require increased frequency resolution by increasing the modulation time $T$ proportionally. Alternatively, if we keep $T$ shorter than the inverse of the energy spacings, 
then the excited states will be effectively blurred into a continuum in energy.  In this regime, experiments can nevertheless access the dynamical structure factor, a physical quantity which encodes universal correlations of primary CFT fields at criticality~\cite{sachdev_quantum_2011}.

In this continuum limit (`large-$L$, short-$T$' limit), the response measured by the modulation-ramp-probe sequence is indeed proportional to the dynamical structure factor, but only at zero temperature and in the low-energy limit.  More precisely, the modulation frequency $f$ must be within the energy range of the first $O(L^2)$ excited states because we ramp to product states at the end of the sequence to probe the transferred population. In the Ising CFT case, this criterion leads to $2\pi f/\Omega \lesssim O(\log^2 L/L)$ (Methods).
We now develop a sequence measuring the dynamical structure factor that does not rely on these conditions and generally applies to any Hamiltonian.

To this end, we develop and show proof-of-principle of an alternative modulation scheme---the \textit{modulation-probe sequence} (Methods)---to measure the system's frequency-dependent response immediately after modulating at the critical point, without an intervening ramp (Fig.~\ref{fig:dynamicSusceptibility}a, top).  
Then, when modulating with and measuring the same operator $\hat K = \sum_j \cos(kj) \hat{o}_j$ ($k$ is the wave vector of the operator $\hat K$ and $\hat{o}_j$ is a local operator), the change $\langle \delta \hat K\rangle = \langle \hat K(T) \rangle -\langle \hat K(0)\rangle$ is proportional to the dynamical structure factor $S(k, \omega)$ at frequency $\omega = 2\pi f$ (Methods). In the experiment, we apply the global modulation $\hat K = \sum_i \hat n_i  \approx \frac{1}{2}\sum_i \hat \varepsilon_i$,
where $\hat{\varepsilon}_{i+\frac{1}{2}} = \hat n_i + \hat{n}_{i+1}$ is the lattice counterpart of the CFT primary field~\cite{slagle_microscopic_2021} $\varepsilon$ with scaling dimension $\Delta_\varepsilon =1$. After modulation, we measure the change in the total number of atoms in the Rydberg state, $\delta n \equiv \langle \delta \hat K\rangle$.

In the peak-resolved limit, this sequence measures the excited state energies, as the response exhibits peaks at the corresponding excitation energies. Using the global modulation, we measure the spectrum of an $L = 19$ array at zero wavevector, $k=0$, and observe distinct spectral peaks that signify excited states (Fig.~\ref{fig:dynamicSusceptibility}a, bottom). However, for a comparable number of projective measurements of $\hat K$ as in the \textit{modulation-ramp-probe} sequence, the signal-to-noise ratio is lower for the \textit{modulation-probe} sequence. This reduction arises because the ground state is no longer an eigenstate of the observable $\hat K$, leading to increased quantum projection noise. Consequently, the \textit{modulation-ramp-probe} sequence is preferable for measuring resolved excitation energies, whereas the \textit{modulation-ramp sequence} is better suited for measuring the universal dynamical structure factor, despite the added complexity of requiring responses measured at two different modulation phases (which we use to suppress higher-order response; Methods) and aforementioned lower signal-to-noise ratio.

In the continuum limit, the sequence measures the dynamical structure factor of the $\varepsilon$ operator of the CFT. The Ising CFT predicts that, at zero temperature, this dynamical structure factor 
\begin{equation}\label{Eq:bulkCFTS}
    S_\varepsilon(k=0,\omega) \sim \omega^{\Delta_\varepsilon-1} = \omega^0
\end{equation}
is a constant at low frequencies, $\omega \ll \Omega$ (Methods). 
Applying this technique to an $L=35$ system tuned to Ising criticality (configuration 1 in the Fig.~\ref{fig:BeyondIsing}a inset), we indeed find an almost-constant response at low frequencies ($\omega \lesssim \Omega$) (Fig.~\ref{fig:dynamicSusceptibility}b), supported also by our tensor network simulations with the time-dependent variational principle (TDVP) method~\cite{hauschild_efficient_2018} (solid line in Fig.~\ref{fig:dynamicSusceptibility}b). 
To evidence that the frequency dependence conforms to a universal scaling function, we numerically simulate the post-modulation responses for several system sizes from $L=35$ to $L=85$ (Fig.~\ref{fig:dynamicSusceptibility}c).  
After rescaling by $(AL)^{-1}$, where $A$ is the modulation amplitude, responses for different system sizes collapse to a universal scaling function that appears to be roughly constant at $\omega/\Omega \lesssim 0.3$---consistent with the CFT prediction (Fig.~\ref{fig:dynamicSusceptibility}c, inset).

Consistency among our experimental results, numerical simulations, and analytical CFT predictions demonstrates that one can obtain the scaling function of the underlying CFT field via measuring the frequency dependence of the system's response to external modulation at the critical point. 
Our numerical simulations also suggest that the modulation-probe sequence is promising for larger systems, which points to future applications of this technique, e.g., combining with the local modulation capabilities to measure the universal $\sigma$-field correlation (we outline the procedure in Methods) and studying many-body systems in higher dimensions.

\vspace{3mm}
\noindent\textbf{Discussion and outlook}

Our experiment harnessed modulation spectroscopy to directly probe excitation spectra of critical Rydberg chains governed by both Ising and TCI universality classes, subject to various tunable fixed-point boundary conditions.  By experimentally implementing finite-size scaling, we recovered the universal dependence of the energy spectrum on system size at criticality.  As a nontrivial consistency check with CFT predictions, we also extracted universal energy ratios among low-energy excited states---which encode linear dispersion, constraints on the admissible operators, as well as scaling dimensions. Furthermore, we carried out a careful theoretical study and directly measured the system’s response at the critical point, allowing us to extract the frequency dependence of the dynamical structure factor governed by CFT field correlations.

The programmability and site-resolved measurement capability of Rydberg atom arrays allows one to probe additional aspects of CFTs as well.  A natural next step is to measure one- and two-point correlators at the fixed points studied here, compare with universal CFT predictions, and quantify discrepancies resulting, e.g., from imperfect ground-state preparation and/or tuning to criticality. By measuring energy and momentum resolved correlation functions, the regime of validity of emergent CFTs can be determined.
Recent theory works propose that, with minimal additional experimental overhead, our setup further enables exploring `measurement-altered quantum criticality'~\cite{AltmanMeasurementLL,JianMeasurementIsing,usmeasurementaltered,EhudMeasurementIsing,patil_highly_2024,liu_boundary_2025}---wherein non-unitary perturbations from measurements qualitatively alter universal CFT predictions~\cite{naus_practical_2025,sarma_probing_2025}. Beyond the Ising and tricritical Ising universality classes explored in this study, similar Rydberg setups could be used to probe other $(1+1)$D CFTs, such as the $c= 4/5$ three-state Potts and the $c=1$ Ashkin-Teller CFT~\cite{maceira_conformal_2022,chepiga_tunable_2024,zhang_probing_2025}. It is also of great theoretical interest to study the preparation of Gibbs states and their properties at  quantum phase transitions~\cite{lloyd_quantum_2025,zhang_quantum_2021}. In addition, these modulation techniques can be extended to probe dynamical properties of quantum critical points in higher dimensions that are difficult to study numerically, such as the dynamical conductivity at the critical point of the 2D ferromagnetic XY model, which admits an AdS/CFT description~\cite{witczak-krempa_dynamics_2014,myers_holographic_2011,chen_universal_2014}.

\FloatBarrier

\bibliography{library_CFT}
\bibliographystyle{apsrev}

\clearpage
\newpage
\setcounter{figure}{0}  
\captionsetup[figure]{labelfont={bf},name={Extended Data Fig.},labelsep=bar,justification=raggedright,font=small}
\section*{Methods}
\noindent\textbf{Experimental setup}

A detailed description of our experimental setup has been given in previous works~\cite{madjarov_high-fidelity_2020,scholl_erasure_2023,Shaw2024}. In short, we trap individual $^{88}\text{Sr}$ atoms in a programmable one-dimensional array of optical tweezers (813 nm) generated by acousto-optic deflectors (AODs). The atoms are initialized in the $5s^2\,^{1}\text{S}_{0}$ state and cooled on the narrow-line red transition $5s^2\,^{1}\text{S}_{0}\leftrightarrow5s5p\,^{3}\text{P}_{1}$ (689 nm) to near their motional ground state. After rearrangement to a defect-free array with the desired atom number, all atoms are driven to the $5s5p\,^{3}\text{P}_{0}$ clock state with a combination of direct $\pi$ pulse (698 nm) and incoherent optical pumping~\cite{scholl_erasure_2023}. Limited by the total number of available tweezers (laser power), we use up to two rounds of dark-state enhanced loading (DSEL) for system sizes of $L=26,27,35$ to ensure high defect-free state preparation fidelity~\cite{Shaw_DSEL_2023}. In our quantum simulator, the metastable clock state $5s5p\,^{3}\text{P}_{0}$ is defined as the ground state $\ket{0}$, which is coupled to the Rydberg state $5s61s\,^{3}\text{S}_{1}\equiv\ket{1}$ with a single-photon transition (317 nm). After evolution under the Hamiltonian in Eq.~\eqref{eq:masterH}, an auto-ionization beam (408 nm) is applied to push out the atoms in the Rydberg state~\cite{madjarov_high-fidelity_2020}. Finally, the remaining atoms are optically pumped out of the clock state and imaged through the $5s^2\,^{1}\text{S}_{0}\leftrightarrow5s5p\,^{1}\text{P}_{1}$ (461 nm) transition~\cite{Covey2019}.

Data for both Ising CFT configurations (configurations (1) and (2) in Fig.~\ref{fig:BeyondIsing}; data shown in Fig.~\ref{fig:isingspectroscopy}-\ref{fig:isingLocal} and Fig.~\ref{fig:BeyondIsing}b) are taken with a Rydberg Rabi frequency $\Omega=2\pi \times 6.0\,\rm MHz$ (with the exception of data presented in Fig.~\ref{fig:isingspectroscopy}d, where $\Omega = 2\pi\times7.5\,\rm{MHz}$) and a next-nearest-neighbor interaction $V_{2}= 2\pi \times 3.06(1)\,\rm MHz$ at a lattice spacing of $a=3.3\,\rm\mu m$ for the Ising transition. For the TCI point (configuration (3)), data are taken with $\Omega= 2\pi \times 5.5\,\rm MHz$, $V_{2}=2\pi \times 8.96(15)\,\rm MHz$ at a spacing of $a=2.8\,\rm\mu m$ for the TCI point. We also experimentally measure the nearest-neighbor interaction $V_{1}=2\pi \times 164.6(9)\,\rm MHz$ at a spacing of $a=3.3\,\rm \mu m$ for the Ising case. The critical detunings for configuration (1), (2), and (3) are $\Delta_c = 2\pi \times 10.2,\, -0.9,\, -8.3\,\text{MHz}$, respectively.

For analysis of the experimental data, we apply a post-selection protocol based on four aspects. First, we post-select on rearrangement success, keeping experimental shots that contain a defect-free array of the correct system size. Second, we post-select based on erasure detection. In the experimental sequence, we perform erasure detection twice: first immediately after the initial state preparation, and second after the Rydberg pulse~\cite{scholl_erasure_2023}. If there are atoms detected in the erasure images, they indicate leakage into the $5s^2\,^{1}\text{S}_{0}$ state, which is outside the qubit subspace. These shots are discarded in the post-selection process. Third, we discard measurement runs where double Rydberg excitations might have occurred. Since we work in the Rydberg blockade regime ($V_1\gg V_2,\Omega,\Delta$), it is unlikely that two nearest-neighboring atoms are both excited to the Rydberg state. We discard all the shots where no atom is detected in consecutive sites after the Rydberg pulse. Note that we do not distinguish the Rydberg occupation from atom loss due to the readout scheme, so this step also filters out shots with loss outside of $5s^2\,^{1}\text{S}_{0}$. Lastly, we post-select based on whether the Rydberg pulse is successfully delivered. The pulse is monitored by a photodiode and recorded on an oscilloscope during the experiment; $\sim0.6\%$ of the shots show no detectable Rydberg pulse due to the arbitrary waveform generator board failing to output the programmed RF waveform, and are thus discarded~\cite{Shaw2024}. For the data shown in the main text, $20-70\%$ of experimental runs are post-selected for analysis.

For the adiabatic sweep, we modulate the Rabi frequency and detuning of the global Rydberg laser beam with AODs. The Rydberg interaction in the system is repulsive ($V_{ij}>0$) at the operating magnetic field $B=70\,\rm G$. In this case, the global detuning is shaped to start from a large negative value and evolve to the critical point under a tangent function in time which ensures that the detuning sweep rate is slower when getting closer to the critical point. Starting from all atoms in state $|0\rangle$, we prepare the ground state at the critical point in this way. To realize an effective attractive interaction ($V_{ij}<0$ for $|i-j|\geq 2$) sector of 
Eq.~\eqref{eq:masterH} in the same system, one can flip the sign of all the terms in the Hamiltonian, which means reversing the sign of detuning relative to the critical point in the ramp ($\Delta-\Delta_c$). Since the nearest-neighbor interaction $V_1$ is much greater than any other energies in the Hamiltonian, the eigenenergies of the Hamiltonian are clustered into sectors separated by $\sim V_1$. Then we prepare the highest-energy state within the blockade-violation-free sector of the full native Hamiltonian, which is the ground state of the blockade-enforced Hamiltonian Eq.~\eqref{eq:masterH} with the sign of interaction and detuning reversed. We call this procedure a backward sweep. Depending on the experiment, the global detuning is then set to either hold at the critical point (for $\sigma$ measurement), or ramp to the $\mathbb{Z}_2$/disordered phase in a symmetric way after modulation (for spectroscopy). These choices, together with the sweep parameters, are optimized such that the system has a minimal number of excitations after the sweep.

\begin{figure*}
    \centering
    \includegraphics[width=0.5\linewidth]{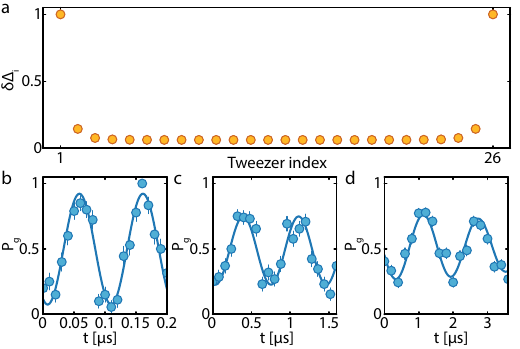}
    \caption{\textbf{Calibration of local detunings of an $L=26$ array.} 
    \textbf{(a)} Normalized local detuning modulation applied to realize (++) boundary condition at TCI.
    \textbf{(b)-(d)} Ramsey oscillation measured at the boundary ($i=1,26$), next-to-boundary ($i=2,25$), and in the bulk ($i=5,11,17,23$). The experimental data is averaged over the sites being measured. Solid lines are fits to a sinusoid with a Gaussian envelop. We extract the local detuning from the oscillation frequency from the fit.
    }
    \label{fig:localdetuningcalib}
\end{figure*}

For experiments that require specific local detuning terms ($\delta\Delta_i$ in the Hamiltonian) during the Rydberg pulse, we apply the same set of tweezers (813 nm) as for trapping. As a reference, for experiments that do not require local detuning terms, the tweezer light is turned off with a fast AOM~\cite{Finkelstein2024} prior to the Rydberg pulse for two reasons: first, intensity noise in the tweezer translates into unwanted detuning noise on the Rydberg-qubit manifold; second, the Rydberg state is anti-trapped in 813-nm light. The tweezer light induces a light shift on the $5s5p\ ^3\text{P}_0 \leftrightarrow 5s61s\  ^3\text{S}_1$ transition that is tunable with control of the local light intensity. We experimentally measure the light shift to be $-3.545(54)\,\rm MHz$ for tweezers with $1.4(2)\,\rm mW$ power per spot and a waist of $0.75(5)\,\rm\mu m$. The time constant of the atom loss due to the anti-trapping of the Rydberg state at this tweezer trap intensity is measured to be $63(12)\,\rm\mu s$. We calibrate the local detuning for each tweezer pattern with a Ramsey sequence before taking the data. Ext. Data Fig.~\ref{fig:localdetuningcalib} shows an example of the calibration in the case of $L=26$ at the tricritical point with the fixed boundary condition.

\vspace{3mm}
\noindent\textbf{Determining the critical point}

To determine the parameters required to prepare our experimental setup along the Ising critical line, we employ exact diagonalization of the effective Hamiltonian in Eq.~\eqref{eq:masterH}. Specifically, we consider the second-order PXP Hamiltonian, first derived in Ref.~\cite{Bluvstein_2021}, using a Schrieffer-Wolf transformation. The resulting Hamiltonian corresponds to Eq.~\eqref{eq:masterH} with $\delta \Delta_i = 0$,
\begin{align*}
V_{ij}=
\begin{cases}
\; \;\;\;V_1 & |i-j|=1\\[4pt]
\dfrac{64V_2}{|i-j|^{6}} & |i-j|\ge 2 
\end{cases}
\end{align*} 
and 

\begin{equation}
\label{eq:2ndorderpxp}
\begin{aligned}
\hat H_2 \;=&\;
-\frac{\Omega^{2}}{4V_1}\Bigg[
\sum_i \Big(
  2\hat n_i -\frac{3}{2}\hat n_{i-1}\hat n_{i+1} \\
&
  +\big(\hat P_{i-1}\hat b_i^{\dagger}\hat b_{i+1}\hat P_{i+2}+\text{h.c.}\big)
\Big)
\Bigg]
\end{aligned}
\end{equation}
where $\hat b_i$ is the hard-core boson operator defined as $\hat b_i =|0\rangle_i\langle1|_i$. 

\begin{figure*}
    \centering
    \includegraphics[width=1\linewidth]{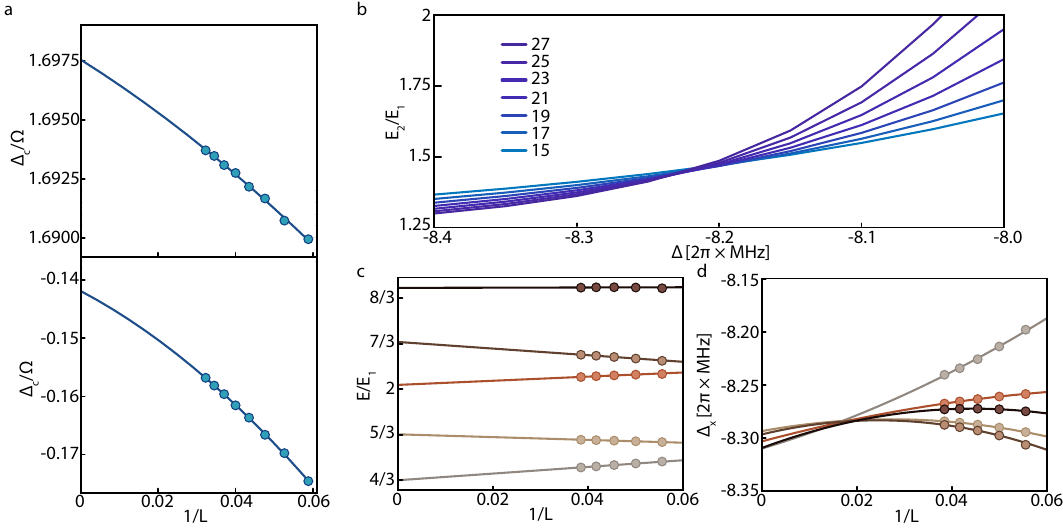}
    \caption{\textbf{Locating the Ising and the tricritical Ising critical point.}  
    \textbf{(a)} Locating the Ising critical point via the finite-size scaling of the curve crossing of $\sigma_{RS}$ for (top) $V_2/\Omega = +0.51$  and (bottom) $V_2/\Omega  = -0.51$. Each data point is the location of $\Delta/\Omega$ where $\sigma_{RS}$ has the same value for chains of length $L-2$ and $L+2$. Using a quadratic fit, we estimate the critical point in the thermodynamic limit at $1/L \to 0$, $\Delta_c(L\to\infty)/\Omega$. For $V_2/\Omega=+0.51,-0.51$, $\Delta_c(L\to\infty)/\Omega \approx 1.6975(4), -0.142(2)$, respectively. 
    \textbf{(b)} Curve crossing of the energy ratio between the second and the first excited state close to the tricritical point. We plot the energy ratio versus detuning $\Delta$ for $\Omega = 2\pi \times 5.5~\rm{MHz}$ and $V_2 = 2\pi \times (-9.0)~\rm{MHz}$ ($V_2/\Omega = -1.63$) for various system sizes $L$. We find the intersection $(\Delta_X, E_2/E_1)$ for two adjacent odd-system sizes $L-1$ and $L+1$ and plot them in \textbf{(c)} and \textbf{(d)}.
    \textbf{(c)} System-size dependence of the energy ratio. We plot the intersecting $E_i/E_1$ for $i \in [2,6]$ and linearly extrapolate to $L\rightarrow\infty$. At this $V_2$, we find best agreements of the first two energy ratio with the TCI spectrum with free boundary condition, $4/3$ and $5/3$, respectively.
    \textbf{(d)} System-size dependence of the intersecting detunings $\Delta_X$. We plot the intersecting $\Delta_X$ for the low-lying energies and quadratically extrapolate to $L\rightarrow\infty$. We find them collapse to $\Delta_c = 2\pi\times (-8.3)~\rm{MHz}$, which we determine to be the critical detuning. 
    }
    \label{fig:isingScalingCollapse}
\end{figure*}

Following the curve-crossing procedure introduced in Ref.~\cite{slagle_microscopic_2021}, we compute the expectation value of $\langle\hat\sigma_{L/2}\rangle$ in the middle of the chain of the ground state for various detunings $\Delta$ while keeping all other Hamiltonian parameters fixed. By performing a finite-size scaling analysis of the crossover point of the rescaled observable $\mathcal{\sigma}_{RS} =\langle \hat \sigma_{L/2} \rangle \cdot \sin(\pi/(L+2))^{-1/8}$ at different values of $L$, we are able to identify the critical point in the thermodynamic limit.  Ext. Data Fig.~\ref{fig:isingScalingCollapse}a illustrates the result of this finite-size scaling analysis for two sets of experimental parameters, $V_2/\Omega = \pm 0.51$.  

Locating the TCI point requires a more costly sweep of two independent Hamiltonian parameters.  We thus utilize a more direct approach to identifying this multicritical point by numerically determining parameters ($V_2$ and $\Delta_c$) that yield a spectrum consistent with TCI CFT predictions. For a fixed interaction strength $V_2$, we calculate the ratio between the energy of excited state $i$ and the energy of the first excited state (relative to the ground state), $E_i/E_1$, for various detunings. As an example, Ext. Data Fig.~\ref{fig:isingScalingCollapse}b shows the energy ratio between the second and the first excited state at $V_2/\Omega = -1.63$ for $L=15$ to $27$.  We determine the values of both $\Delta_X$ and $E_i/E_1$ from where the curves for each pair of system sizes $L-1$ and $L+1$ intersect and then extrapolate these values to the thermodynamic limit $1/L \rightarrow 0$. We then vary $V_2$ compare the extrapolated energy ratio with the free-boundary TCI spectrum 3:4:5:6:7:8:$\cdots$ and determine the critical interaction strength $V_2$ (Ext. Data Fig.~\ref{fig:isingScalingCollapse}c). At $V_2/\Omega = -1.63$, $E_2/E_1$ and $E_3/E_1$ best fit the predicted ratios of $4/3$ and $5/3$, respectively. Under this interaction strength, the detunings $\Delta_X$ for different energy levels also converge to the same value in the thermodynamic limit, where we find $\Delta_c/\Omega = -1.51$ (Ext. Data Fig.~\ref{fig:isingScalingCollapse}d).

\vspace{3mm}
\noindent\textbf{Coherent control of many-body states}

In the main text, we use modulation techniques to measure the excited state spectrum. When the system is prepared in the ground state and the modulation is resonant with the energy difference between the ground state and a chosen excited state, then the population is transferred to that excited state. As we will show, the population transfer is a coherent oscillation (Rabi oscillation) between the ground state and the excited state. Here, we isolate the ground state and the first excited state for a Hamiltonian slightly away from the critical point and apply modulation to observe the Rabi oscillation. Furthermore, we prepare an equal superposition between the ground state and the first excited state and measure the coherence time via Ramsey interferometry.

We first outline the scheme of driving a coherent oscillation between two many-body states. Suppose that the system is initialized in the ground state $\ket{g}$ of the Hamiltonian $\hat H$, which can be written in its diagonal form:
\begin{equation}
    \hat H = E_g \ketbra{g}+ (E_g+E_1) \ketbra{e_1} + \cdots
\end{equation}
where the ellipsis contains matrix elements among all other states orthogonal to both $\ket{g}$ and the first excited state $\ket{e_1}$. Now we perturbatively modulate the system with $\delta \hat H = A \cos(E_1 t + \varphi) \hat K$. Now we change into a rotating frame with $\hat U = \sum_j e^{-iE_jt} \ketbra{e_j}$, in this frame, the effective Hamiltonian is
\begin{equation}
\begin{split}
    \hat H' &= \hat U^\dagger (\hat H + \delta \hat H) \hat U\\
    &= E_g +  \dfrac{A}{2} \left(\bra{g}\hat K\ket{e_1} e^{-i\varphi}\ket{g}\bra{e_1} + h.c.\right) + \cdots.
\end{split}
\end{equation}
When the first excited state is non-degenerate and there is not a state whose energy is $E_1$ higher than the first excited state, the ellipsis only contains irrelevant fast-oscillating terms that can be dropped out. Then, the dynamics is restricted to a two-level system with a many-body Rabi frequency $\Omega_\text{MB} = A \left|\bra{g}\hat K\ket{e_1}\right|$.

\begin{figure*}[]
    \centering
    \includegraphics[width=0.5\linewidth]{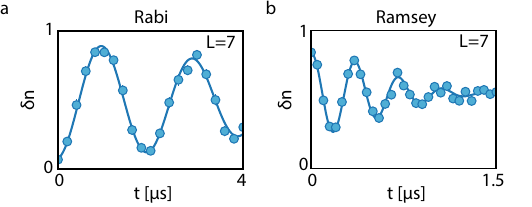}
    \caption{\textbf{Many-body Rabi oscillation and coherence on a 7-atom array.} 
    \textbf{(a)} Rabi oscillation between the ground and the first excited state. We modulate the system with the frequency being the energy difference between the two states and measure the final change of ground state population, $\delta n$, after ramping into the $\mathbb{Z}_2$ phase. When $\delta n = 0 / 1$, it indicates that the system is in its ground / excited state. The solid line is a fit to a decaying sinusoidal oscillation.
    \textbf{(b)} Ramsey interrogation of the energy gap. Using the many-body Rabi frequency determined from \textbf{(a)}, we apply two $\pi/2$ rotations, separated by a time $t$. After the Ramsey sequence, the system is ramped into $\mathbb{Z}_2$ phase for readout. We observe a damped oscillation, whose oscillation frequency is consistent with the energy gap. The solid line is a fit to a decaying sinusoidal oscillation.
    }
    \label{fig:RabiRamsey}
\end{figure*}

Here we describe the experimental sequence to measure the Rabi oscillation on a 7-atom array. Experiments begin with all atoms initialized in the electronic ground state $\ket{0}$, which is approximately the many-body ground state in the disordered phase. Then, we adiabatically ramp the detuning to $\Delta = 2\pi \times 8.8~\mathrm{MHz}$ to prepare the many-body ground state at this detuning. This is slightly away from the critical point, because at the critical point, transition from the first excited state to a higher excited state is nearly resonant with $E_1$. Then, we apply a global detuning modulation with frequency $\omega = 2\pi \times 2.83~\text{MHz}$ (measured by the modulation spectroscopy) for a variable time $t$. Finally, we adiabatically ramp into the $\mathbb{Z}_2$ ordered phase and read out the total number of atoms in the ground state $\hat O' = \sum_i (1-\hat{n}_i)$. The ground state and the first excited state are adiabatically transferred to $\ket{1010101}$ and $\ket{1001001}$, respectively. Hence, we define $\delta n \equiv \langle \hat O'\rangle-3$. When the final state is the ground state, $\delta n = 0$, and when the final state is the first excited state, $\delta n=1$. We observe an oscillation of $\delta n$ when we vary the modulation time $t$ (Ext. Data Fig.~\ref{fig:RabiRamsey}a). We fit the data to a Gaussian decaying oscillation and find an oscillation frequency of $0.506(5)~\text{MHz}$, an initial contrast of $0.83(3)$, and an $1/e$ coherence time of $5.5(5)~\mu\text{s}$.

Next, we perform an accurate measurement of the first excited energy $E_1$ and characterize the many-body phase coherence with the Ramsey interferometry sequence. After preparing the ground state, we apply a $\pi/2$ pulse by driving a quarter cycle of Rabi oscillation. This prepares the system in $(\ket{g} + \ket{e_1})/\sqrt{2}$. Then, we turn off the modulation and let the system evolve under $\hat H$ for time $t$. The state after evolving under $\hat H$ for time $t$ is $\ket{\psi(t)} = \left(\ket{g} + e^{-iE_1 t}\ket{e_1} \right)/\sqrt{2}$. We subsequently apply another $\pi/2$ pulse with a fixed initial phase. In the absence of any decoherence, the population in the excited state is $P_e(t) = \left[1+\cos(E_1t)\right]/2$. In the experiment, we observe a decaying oscillation with an initial contrast of $0.58(2)$ and an oscillation frequency $2.81(3)~\text{MHz}$ (Ext. Data Fig.~\ref{fig:RabiRamsey}b), which is consistent with the first excited state energy $E_1 = h \times 2.83(4)~\text{MHz}$ obtained from the modulation spectroscopy. We fit the decay to a Gaussian function and find a coherence time of $0.75(4)~\mu\text{s}$. We postulate that the dominant factors of the contrast decay of both the Rabi oscillation and the Ramsey oscillation are the intensity noise and the frequency noise of the Rydberg laser.

\vspace{3mm}
\noindent\textbf{Modulation spectroscopy}

In the main text, we implemented two types of modulation spectroscopy sequences: the \textit{modulation-ramp-probe sequence} which we primarily use to measure the excited state energies of the Hamiltonian, and the \textit{modulation-probe sequence} which we primarily use to measure the dynamical structure factor of the Hamiltonian.  As the names suggest, the former sequence measures the Rydberg occupation numbers only after ramping to the CDW or trivial phase, while the latter sequence performs the measurements directly on the modulated state.  
Both sequences can also operate in two different regimes: when probing the state-resolved spectrum, the modulation time $T$ is long enough to resolve the individual energy levels (\textit{peak-resolved regime}, $T\gg E_\Delta^{-1}$, where $E_\Delta$ is the energy gap that is a characteristic energy scale), whereas when measuring the dynamical structure factor, we use a short modulation time such that excited states will be effectively blurred into a continuum in energy (\textit{continuum regime}, $T \ll E_{\Delta}^{-1}$). Although either sequence can measure both the excitation spectra and the dynamical structure factor, depending on the task, one of the sequences tends to be preferable over the other one for reasons that we will explain in this section.

In what follows we first use perturbation theory to derive signals of both modulation spectroscopy sequences, and then discuss the relation of these signals to the \textit{dynamical structure factor} introduced later in this section.\\

\noindent\textit{Modulation-ramp-probe sequence}
\vspace{3mm}

The modulation-ramp-probe sequence reveals the energy differences between the excited states and the ground state of a many-body Hamiltonian $\hat H$ along with the transition strengths between these states. The key idea behind this sequence is to apply a sinusoidal modulation and measure the quadratic response of an observable $\hat{O}$ that is diagonal in the eigenbasis of $\hat H$. There will be a significant change in $\hat O$ when the modulation frequency equals the energy difference of an excited state and the ground state. In the experiment, we only measure the local Rydberg excitations $\{\hat n_i\}$. However, linear combinations of these operators are generally not simultaneously diagonalizable with $\hat H$ that we want to probe. Hence, we adiabatically ramp to a Hamiltonian $\hat H'$ where $\{\hat n_i\}$ is simultaneously diagonalizable with $\hat H'$ and perform measurements. In the experiment, we choose to measure the total number of atoms in the ground state, $\hat O' = \sum_i (1-\hat n_i)$, that is simultaneously diagonalizable with the Hamiltonian $\hat H'$ describing the system deep in the trivial phase or deep in the CDW phase. As we will show later in this section, this is equivalent to measuring an observable $\hat O$, that is diagonalized in the eigenbasis of $\hat H$.

We first calculate the post-modulation signal $\delta \langle \hat O \rangle$, which is the difference between $\langle \hat O \rangle$ with and without modulation. Suppose that we prepare the ground state $\ket g$ of $\hat{H}$ and apply a modulation $\delta \hat H = A(t) \hat K$.
Here, we choose the modulation profile $A(t)$ to be a sinusoidal pulse superimposed by a Gaussian envelope:
\begin{equation}\label{eq:modulationAmplitude}
    A(t) = A\, e^{-\left(\frac{t-T/2}{w}\right)^2} \cos(\omega t+\varphi),\ t\in [0, T],
\end{equation}
where $A$ is the modulation amplitude and $\varphi$ is a random phase. Importantly, since $\hat O$ is diagonalized in the eigenbasis of $\hat H$, the change $\delta \langle \hat O \rangle$ is proportional to the population in excited states. Therefore, the leading-order term of $\delta \langle\hat O \rangle$ is quadratic in $A$. We use second-order time-dependent perturbation theory to calculate $\delta \langle\hat O \rangle$:
\begin{widetext}
\begin{equation}\label{eq:quadraticResponseFull}
        \delta \langle \hat O \rangle = \dfrac{\pi w^2 A^2}{4} \sum_{e}  (O_{ee} - O_{gg}) |K_{ge}|^2  
        \left[e^{-\frac{w^2}{2}(\omega + \Delta E_{eg})^2} + e^{-\frac{w^2}{2}(\omega - \Delta E_{eg})^2} + 2 e^{-\frac{w^2}{2}(\omega^2 + \Delta E_{eg}^2)} \cos(2\varphi)\right],
\end{equation}
\end{widetext}
where $\Delta E_{eg} = E_e-E_g$ is the energy difference between the excited state and the ground state, $O_{ee}\equiv\bra{e}\hat O \ket{e}$ and $O_{gg}\equiv\bra{g}\hat O \ket{g}$ are diagonal matrix elements of $\hat O$, and $|K_{ge}|^2 \equiv\left| \bra{g}\hat K \ket{e} \right|^2$ is the off-diagonal transition matrix element between these two states. 

To resolve the energy of each individual state, we work in the peak-resolved regime, where the frequency width of each excited-state response ($\Delta \omega = \sqrt{2 \ln 2} w^{-1}$) is narrower than the spacing between energy levels. In this regime, the first and the third terms of Eq.~\eqref{eq:quadraticResponseFull} are exponentially small; hence we can neglect them. The change $\delta \langle \hat O \rangle$ then reduces to a sum of Gaussian functions, each of which is peaked at $E_e-E_g$:
\begin{equation}\label{eq:response}
    \delta \langle \hat O \rangle = \dfrac{\pi w^2 A^2}{4} \sum_{e}  (O_{ee} - O_{gg}) |K_{ge}|^2  
        e^{-\frac{w^2}{2}(\omega - \Delta E_{eg})^2}.
\end{equation}
To measure the excited state spectrum, we should choose $\hat O$ such that $O_{gg}$ is distinct from all excited states' $O_{ee}$.

\begin{figure*}[]
    \centering
    \includegraphics[width=1\linewidth]{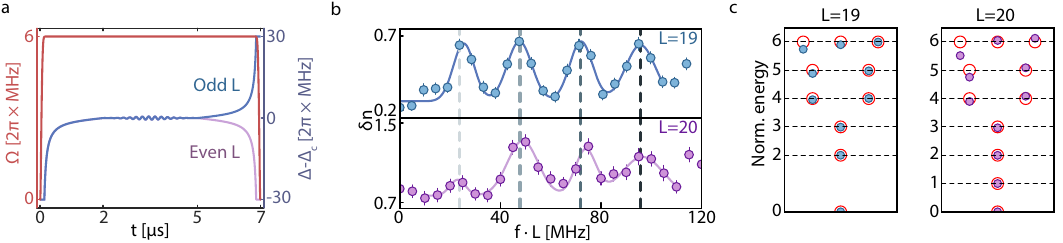}
    \caption{\textbf{Spectra of odd-length and even-length arrays at Ising criticality.}
    \textbf{(a)} Exemplary Rabi frequency and detuning profiles of modulation-ramp-probe sequences for odd- and even-length chains.
    \textbf{(b)} Experimental measurement of spectrum of $L=20$ array, compared with the spectrum of an $L=19$ array. The $x$-axis is rescaled with the system size $L$. After the rescaling, the peak positions collapse to the same non-universal values, with a universal ratio of 2:4:6:8. Solid lines are multi-Gaussian fits to the data and indicate fit ranges.
    \textbf{(c)} Numerical calculation of excited state energies. We numerically calculate the eigenenergies of an $L=19$ and $L=20$ array, tuned to Ising criticality. Numerically calculated spectra (blue and purple circles) are normalized by the energy difference between the first excited and the ground state and are compared against Ising CFT spectra (red open circles). There is one additional state with normalized energy of $1$ for an even array, and the rest of the spectra remain the same.
    }
    \label{fig:EvenN}
\end{figure*}

We now describe the experimental sequence that aligns with the formalism outlined above. Experiments begin with preparing the many-body ground state of the Hamiltonian $\hat{H}$ using adiabatic state preparation, as described in the previous section. The sweep profile is illustrated in Ext. Data Fig.~\ref{fig:EvenN}a. Next we apply a modulation, also as described before. Eventually, to perform the measurement, we adiabatically ramp the detuning into the disordered / $\mathbb{Z}_2$ phase and measure the number of atoms in the Rydberg state / ground state, $\hat{O}' = \sum_i \hat n_i$ or $\sum_i (1-\hat n_i)$, respectively. We will show that this is effectively measuring an observable $\hat O$ that is diagonal in the eigenbasis of the critical Hamiltonian and that $O_{gg}$ is distinct from all $O_{ee}$. The change in $\hat{O}'$, comparing with the expectation value of the ground state, is defined as $\delta n$. Assuming this fact, peaks in the measured signal indicate population transferred from the ground state to excited states. This allows us to fit the experimental data and extract the peak centers, which we determine as excited state energies.

We now show that the measurement protocol is equivalent to measuring an operator
$\hat O$ immediately after the modulation. To this end, we first identify the
structure of the operator $\hat O'$. Deep in the disordered phase or in the $\mathbb{Z}_2$-ordered phase
($|\Delta|\gg |V_2|,|\Omega|$), the Hamiltonian can be approximated as
\begin{equation}
    \hat H' \simeq -\Delta \sum_i \hat n_i = |\Delta| \hat O' +E_0,
\end{equation}
where $E_0$ is a constant energy. In this regime, $\hat O'$ is diagonal in the eigenbasis of $\hat H'$. In the
disordered phase, the ground state is unique, $\ket{0}^{\otimes L}$, while the
first excited manifold consists of single-excitation states
$\ket{0\cdots010\cdots0}$ with degeneracy $L$. Restricting to the low-energy subspace spanned by the ground state $\ket{g'}$ and
the single-excitation states $\ket{e'}$, the operator $\hat O'$ takes the form
$ \hat O' = C + \sum_{e' \neq g'} \ketbra{e'}$, where $C=\bra{g'}\hat O'\ket{g'}$ is a constant depending only on the system size
$L$ and on the phase reached by the ramp. It follows that, in the low-energy
limit, measuring $\hat O'$ is equivalent to measuring $\hat O \equiv C + \sum_{e \neq g} \ketbra{e}$, where $\ket{e}$ denote the excited states of the Hamiltonian $\hat H$.

We model the adiabatic ramp after the detuning as a unitary $\hat U$. By the adiabatic theorem, $\hat U$ maps the ground state $\ket{g}$ of $\hat H$ to the ground state $\ket{g'}$ of $\hat{H}'$ and maps a low-energy excited state $\ket{e}$ to a state $\ket{e'}$ in the first degenerate excited subspace. Therefore, the observable $\hat{O}$, in the Heisenberg picture, becomes
\begin{equation}
\begin{split}
    \hat O_H &= \hat U \hat O \hat U^\dagger\\
    &= C + \sum_e U \ketbra{e} U^\dagger \\ 
    &= C+\sum_{e'} \ketbra{e'}\\
    &= \hat{O}'.
\end{split}
\end{equation}
This further implies that measuring $\hat{O}'$ after ramping into the disordered phase is equivalent to measuring $\hat O$ right after the modulation:
\begin{equation}
    \bra{\psi'}\hat{O}'\ket{\psi'} = \bra{\psi} \hat U^\dagger (\hat U \hat O \hat U^\dagger) \hat U \ket{\psi} = \bra{\psi} \hat O \ket{\psi}.
\end{equation}
In the disordered phase, $O_{gg}$ is indeed distinct from all $O_{ee}$ of excited states because the ground state of the disordered phase is unique. 

In practice, ramping to the $\mathbb{Z}_2$ phase for measurement on an odd chain (Ext. Data Fig.~\ref{fig:EvenN}a) is more preferable for two reasons. First, the ground state of the $\mathbb{Z}_2$ phase for an odd chain is still unique ($\ket{1010\cdots 0101}$), and the first excited energy level contains all states with the creation of a pair of domain walls, with a degeneracy of $(L+3)(L+1)/8$, higher than the degeneracy with ramping into the disordered phase. This increases the energy range where Eq.~\eqref{eq:peakHeight} holds. Second, when probing the critical spectra, since the energy gap reaches its minimum in the disordered phase for a finite system size, ramping into the $\mathbb{Z}_2$ phase afterwards avoids passing through the small gap twice; this reduces the number of excitations created by the non-adiabaticity of the ramp and hence enhances the contrast of the signal. Moreover, since $O_{ee} - O_{gg} = 1$, independent of the phase we ramp to, we can drop the $O$-dependent factor in Eq.~\eqref{eq:response}, which reduces to Eq.~\eqref{eq:peakHeight} in the main text. Hence, we can interpret the height of a spectral peak as the transition strength $|K_{ge}|^2$.

For an even chain, since the ground state of $\mathbb{Z}_2$ phase is degenerate, we have to ramp back into the disordered phase to probe the excited states (Ext. Data Fig.~\ref{fig:EvenN}a). The result for an $L=20$ chain is presented in Ext. Data Fig.~\ref{fig:EvenN}b. After multiplying the modulation frequency $\omega = 2\pi f$ by the system size $L$, we find the even-parity peak positions of an odd chain $L=19$ and the even chain $L=20$ collapse to a same set of values. This is due to CFT predictions of odd chains and even chains being the same,  except for an additional odd-parity state with normalized energy of 1 that only arises for a chain with an even number of sites (Ext. Data Fig.~\ref{fig:EvenN}c).

\vspace{3mm}
\noindent\textit{Modulation-probe sequence}
\vspace{3mm}

The modulation-probe sequence also utilizes the modulation technique that reveals the structure of the Hamiltonian $\hat H$. The idea is to measure the linear response of an operator $\hat Q$, not diagonalized in the eigenbasis of $\hat H$ but directly measurable at $\hat H$, under a modulation $\hat K$. In the peak-resolved regime, it also reveals the spectrum of $\hat H$; in the continuum limit, the measurement signal is related to the dynamical structure factor.

Similar to the previous section, we calculate the post-modulation signal $\delta \langle \hat Q\rangle$. We choose to modulate the system with $\delta \hat H = A(t) \hat K$, where the modulation profile $A(t)$ is a product of a normalized envelop function $f(t)$ (\textit{e.g.,} a Gaussian function or a square pulse) and a sinusoidal oscillation, scaled by an overall amplitude $A$:
\begin{equation}
    A(t) = A\, f(t) \cos(\omega t+\varphi), \ t\in [0, T].
\end{equation}
Assuming we prepare the ground state, the modulation creates a superposition of the ground state and excited states. Hence, the leading-order contribution to $\delta \langle \hat Q\rangle$ is from the off-diagonal matrix elements $K_{ge}Q_{eg}$. and the leading-order response is linear in $A$. More generally, if we assume that the initial state is a Gibbs state with inverse temperature $\beta$, the change in $\langle \hat Q \rangle$, to the linear order in $A$, is
\begin{widetext}
\begin{equation}\label{eq:LinearResponse}
    \delta \langle \hat{Q}\rangle (\omega) = \dfrac{\sqrt{2\pi}A}{\mathcal{Z}} \sum_{m,n} (e^{-\beta E_m} - e^{-\beta E_n})\mathrm{Im}\left[Q_{mn} K_{nm} F\left(\omega-(E_n - E_m)\right) e^{i(E_m-E_n)T}e^{-i\varphi} \right],
\end{equation}
\end{widetext}
where the sum is performed over all pairs of eigenstates $(\ket{m}, \ket{n})$ of $\hat H$, and $F(\omega)$ is the Fourier transform of the of the temporal envelop function $f(t)$. For common envelop functions (square pulse, Gaussian function, etc.), $|F(\omega)|$ peaks at $\omega=0$. Hence, this also results in strong responses when the modulation frequency $\omega$ is resonant with the energy difference $E_n-E_m$ between a pair of states in the peak-resolved regime. 

Although the modulation-probe sequence can also measure the excitation energies, the observable $\hat Q$ is not simultaneously diagonalizable with $\hat H$, resulting in larger projection noise since the initial state is not an eigenstate of $\hat Q$ (Fig.~\ref{fig:dynamicSusceptibility}a). On the contrary, the observable in the modulation-ramp-probe sequence is, by construction, diagonalized in the eigenbasis of $\hat H$; consequently, it is preferable to use the modulation-ramp-probe sequence to measure the excitation energies.

When the modulation time is fixed while the system size becomes larger, energy spacings between excited states become smaller compared to the width of $F(\omega)$. Here, we consider the other extreme where there are many states within the width of $F(\omega)$, i.e.,  the continuum limit. In this limit, at frequency $\omega$, we need to sum over the contribution from a continuum of states. The response of $\langle\hat Q \rangle$ is then
\begin{widetext}
    \begin{equation}\label{eq:LinearResponseCont}
        \delta \langle \hat Q \rangle^{\rm cont.}(\omega)  = \dfrac{\pi Af(T^-)}{ \mathcal{Z}}(1-e^{-\beta \omega})\sum_{m,n} {e^{-\beta E_m}} \mathrm{Im}\left[Q_{mn}K_{nm}e^{-i(\omega T+\varphi)}\right] \delta(\omega-(E_n-E_m)),
    \end{equation}
\end{widetext}
where $f(T^-)$ is the left limit of the temporal envelop function at time $T$. The response $\delta \langle \hat Q \rangle^{\rm cont.}$ is now written explicitly in the form of Fermi’s golden rule and can be related to the dynamical structure factor of the system.

Now we describe an experimental protocol to measure the linear response as in Eqs.~\eqref{eq:LinearResponse} and~\eqref{eq:LinearResponseCont}. As in the modulation-ramp-probe sequence, we start with preparing the ground state in the disordered phase and adiabatically ramp the system to the critical point, which ideally initializes the system in the ground state of the critical Hamiltonian. Then, we perform a modulation $\delta \hat H$, followed by a measurement of $\hat Q$, which is a function of local $\hat n_i$'s, immediately after the modulation. The linear response depends on the modulation's final phase $\varphi + \omega T$, so in principle one would need to vary $\varphi$ to obtain both the amplitude and the phase information (or real and imaginary part) of $Q_{mn}K_{nm}$. 

To obtain the linear response of $\delta \langle \hat Q\rangle$, one needs to choose a small $A$ such that higher-order responses (which we did not include in Eqs.~\eqref{eq:LinearResponse} and~\eqref{eq:LinearResponseCont}) are smaller than the linear-order term. However, such an $A$ could be too small for the experiment to acquire a sizable signal of the linear response. Here, we utilize the following protocol to eliminate all even-order responses (dominated by the second-order response). Notice that in Eq.~\eqref{eq:LinearResponse} and~\eqref{eq:LinearResponseCont}, when we change the phase $\varphi$ to $\varphi + \pi$, the responses will acquire a minus sign. Yet, the $\varphi$-dependence of the second-order response is a polynomial of $e^{i2\varphi}$, such that changing $\varphi$ to $\varphi + \pi$ does not affect the second-order response (and all even-order responses). Therefore, when we subtract the two measured signals with $\varphi$ and $\varphi + \pi$, all even-order responses cancel, and the linear response survives. When $\hat Q = \hat K$, which is relevant for the experiment, $\varphi$ is chosen to be $\pm \pi/2 - \omega T$ to maximize the response.

\vspace{3mm}

\noindent\textit{Dynamical structure factor}
\vspace{3mm}

The dynamical structure factor is the Fourier transform of the spatial and temporal correlation in a system, which encodes the momentum and frequency information of excitations in the system. Here, we first define the momentum- and frequency-dependent dynamical structure factor of a local ``field'' operator $\hat o_i$:
\begin{widetext}
    \begin{equation}\label{eq:dynamicalStructureFactor}
    \begin{split}
    S(k, \omega) &= \dfrac{1}{L \mathcal{Z}} \sum_{j,l} \int_{-\infty}^{\infty}{\rm d}t\, e^{-ik(j-l)+i\omega t} \mathrm{Tr}[e^{-\beta H}\hat{o}_j(t) \hat{o}_l (0)]\\
    & = \dfrac{1}{L\mathcal{Z}} \sum_{m,n}\sum_{j,l}e^{-\beta E_m} e^{-ik(j-l)}   (o_j)_{mn} (o_l)_{nm}  \delta(\omega-(E_n-E_m)).
    \end{split}
    \end{equation}
\end{widetext}
We can relate Eq.~\eqref{eq:LinearResponseCont} and Eq.~\eqref{eq:dynamicalStructureFactor} by plugging in $\hat Q=\sum_j e^{-ikj}\hat o_j$ and $\hat K = \hat Q^\dagger$. Further, we choose the modulation final phase $\omega T + \varphi = - \pi/2$ to maximize the linear response. Then, we find 
\begin{equation}\label{eq:dynamicalStructureFactorandResponse}
\begin{split}
    S(k, \omega) &= \dfrac{1}{L\mathcal{Z}} \sum_{m,n}e^{-\beta E_m} Q_{mn} K_{nm}  \delta(\omega-(E_n-E_m))\\
    &= \dfrac{1}{\pi f(T^-)AL(1-e^{-\beta \omega})} \delta\langle \hat Q\rangle^{\rm cont.}(\omega),
\end{split}
\end{equation}
suggesting that in the continuum limit, the measured linear response from the modulation-probe sequence is proportional to the dynamical structure factor. 

Finally, we demonstrate that not only the modulation-probe but also the modulation-ramp-probe sequence measures the universal dynamical structure factor at low frequencies in the continuum limit, under certain conditions. Noting that since the first excited state after ramping is highly-degenerate, if we take the low-energy limit of Eq.~\eqref{eq:response}, where we only perform the sum over excited states in the first excited degenerate state manifold, then we can replace $O_{ee}-O_{gg}$ with 1, and the formula is reduced to, in the continuum limit, 
\begin{equation}
    \delta\langle \hat{O} \rangle^{\rm cont.} = \dfrac{\sqrt{2\pi^3} w A^2}{4} \sum_e |K_{ge}|^2 \delta(\omega-(E_e-E_g)).
\end{equation}
When we choose $\hat{K} = \sum_j e^{-ikj} \hat o_j$, then $\delta \langle \hat O \rangle^{\rm cont.} \propto S(k, \omega)$ of zero temperature.

These results imply that when we approach larger system sizes, where states occupy an approximate continuum of energies and resolving individual excited state energies becomes less practical, we can migrate both modulation techniques into measuring the physically-relevant dynamical structure factor, which we will show conforms to a universal CFT prediction at the critical point. 
The modulation-ramp-probe sequence, however, only reproduces $S$ at zero temperature (which is in general difficult to access with an analog quantum simulator) and in the low-energy limit; specifically, in the Ising CFT case, the response is proportional to the dynamical structure factor only at $\omega/\Omega \lesssim O(\log^2 L/L)$. Therefore, we will implement the modulation-probe sequence to measure the dynamical structure factor, as it has less restrictions.

\vspace{3mm}
\noindent\textbf{Analysis of measured spectra}

We use the modulation-ramp-probe sequence to extract the excited state energies, relative to the ground state energy. To this end, we fit the measured spectra to some trial function. In the data presented in Figs.~\ref{fig:isingspectroscopy}, \ref{fig:isingscaling}, \ref{fig:isingLocal}, and \ref{fig:BeyondIsing}, we fit the data to a sum of Gaussian functions,
\begin{equation}
    \delta n(f)=a_0+\sum_i a_i e^{-\left(\frac{f-E_i/h}{w_i}\right)^2}
\end{equation}
and refer to the centers $E_i>0$ as excited state energies. In Fig.~\ref{fig:TCI}c, we choose a shorter modulation time and fit the spectra to
\begin{widetext}
\begin{equation}
    \delta n(f)=a_0+\sum_i a_i \left[e^{-\left(\frac{f-E_i/h}{w_i}\right)^2} + e^{-\left(\frac{f+E_i/h}{w_i}\right)^2} + 2 e^{-\frac{f^2+(E_i/h)^2}{w_i^2}} \cos(2\varphi_f)\right]
\end{equation}
\end{widetext}
where $\varphi_f$ is the phase that we choose during the modulation.

We find optimal fitting parameters via minimizing $\chi^2=\sum_j [\delta n_\text{exp}(f_j)-\delta n_\text{model}(f_j)]^2/\sigma_j^2$, where $\sigma_j$ is the experimental statistical uncertainty for data taken at $f_j$. For the extraction of the fitting uncertainties, we perform a bootstrap method on the data: we resample the data with a normal distribution $\mathcal{N}(\delta n_\text{exp}(f_j), \sigma_j^2)$ and fit the generated data set with the model. The uncertainty of the fit parameter is quoted as the standard deviation of the optimal fit parameters of resampled data sets.

With this fitting method, we test the hypothesis that the critical spectrum, as presented in Fig.~\ref{fig:isingScalingCollapse}c, is described by the Ising CFT spectrum. We calculate the chi-square, $\chi^2 = \sum_i (f_j^\text{fit}-f_j^\text{model})^2/(\sigma_j^\text{fit})^2$, for the fit frequencies, with a model prediction without free parameters. When we include all data with $L \ge 19$, we find all data agree with the model prediction within $3\sigma$ with a reduced chi-square of $\chi^2/\nu = 0.98$. This shows that critical spectra for system sizes $L \ge 19$ are consistent with the Ising CFT spectrum, indicating that our system is described by an Ising CFT.

\vspace{3mm}

\noindent\textbf{Low-energy Ising and TCI excitation spectra}

Along the second-order Ising phase boundary, low-energy excitations are created by emergent right- and left-moving Majorana fermions $\gamma_{R/L}(x)$, where $x$ is a coarse-grained position.  Microscopically, the fermions arise from a product of the CDW order parameter and a `disorder parameter' corresponding to a non-local string operator that creates a CDW domain wall \cite{slagle_microscopic_2021}. The effective low-energy Hamiltonian for an open length-$L$ Rydberg chain is given by
\begin{equation}
    \mathcal{H}_{\rm Ising} = \int_{-L/2}^{L/2}dx(-i\hbar v \gamma_R \partial_x \gamma_R + i\hbar v \gamma_L \partial_x \gamma_L)
    \label{HIsing}
\end{equation}
with $v$ a non-universal velocity.  The microscopic Rydberg chain Hamiltonian would additionally yield higher-derivative and interaction terms that we did not include in $\mathcal{H}_{\rm Ising}$, but they are irrelevant and can thus be neglected when discussing low-energy excitations.  For convenience in addressing reflection symmetry below, we defined the chain to live on the interval from $x = -L/2$ to $+L/2$.  When a right-mover hits the boundary it must backscatter into a left-mover (and vice versa); hence $\gamma_{R/L}$ satisfy certain relations at $\pm L/2$ that are tightly constrained by Hermiticity of $\gamma_{R/L}$ and the need for a countable Hilbert space.  We adopt a convention such that
\begin{equation}
    \gamma_R(-L/2) = \gamma_L(-L/2),~~~~~ \gamma_R(L/2) = -\gamma_L(L/2).
    \label{BCs}
\end{equation}  The minus sign in the last equation is crucial: Had we taken $\gamma_R = \gamma_L$ at both endpoints, the spectrum would feature a single Majorana zero mode---which does not yield a sensible Hilbert space since Majorana zero modes invariably come in pairs.  

Diagonalizing Eq.~\eqref{HIsing} yields (up to a constant)
\begin{equation}
    \mathcal{H}_{\rm Ising} = \sum_{k_n>0} \hbar v k_n \Gamma_{k_n}^\dagger \Gamma_{k_n},
\end{equation}
where $\Gamma_{k_n}^\dagger$ creates an excitation with momentum 
\begin{equation}
    k_n = \frac{\pi}{L}(n+1/2),~~~n \in \mathbb{Z}
\end{equation}
and energy $\hbar v k_n$.
The momentum quantization condition above can be efficiently recovered by combining the right- and left-movers into a single \emph{chiral} fermion living on a perimeter of length $2L$ with \emph{anti-periodic} boundary conditions by virtue of Eq.~\eqref{BCs}. 

To assess reflection properties of the energy eigenstates, we decompose $\Gamma_{k_n}$ in terms of $\gamma_{R/L}$ via
\begin{equation}
  \Gamma_{k_n} = \frac{1}{\sqrt{2L}}\int_{-L/2}^{L/2} dx[e^{-i k_n x}\gamma_R -i(-1)^{n+1} e^{i k_n x}\gamma_L].
  \label{eq:GammaOps}
\end{equation}
Reflection---which swaps right- and left-movers---sends  
\begin{align}
    R_x: & ~ \gamma_R(x) \rightarrow -i(-1)^{L+1} \gamma_L(-x) G
    \nonumber \\
    &~\gamma_L(x) \rightarrow i(-1)^{L+1}\gamma_R(-x)G.
    \label{eq:gammaRx}
\end{align}  
The relative sign difference in the top and bottom transformation is necessary to maintain invariance of Eq.~\eqref{BCs} under reflection. We use numerically calculated reflection eigenvalues to fix the convention so the top transformation possesses the additional minus sign. The $(-1)^{L+1}$ factors arise because reflections are bond-centered for even $L$ but site-centered for odd $L$.  Under the appropriate reflection, the coarse-grained CDW order parameter transforms as\cite{slagle_microscopic_2021} $\sigma(x) \rightarrow (-1)^{L+1} \sigma(-x)$.  Fermions---which are again products of order and disorder parameters---inherit the sign above as incorporated in Eq.~\eqref{eq:gammaRx}.  The operators $G$ account for the fact that reflection switches the orientation of the non-local string operator in the microscopic definition of the fermions; $G$, which counts the global fermion parity, switches the orientation back.  Finally, $G$ and $\gamma_{R,L}$ anticommute, and hence the factors of $i$ are required to maintain Hermiticity of the Majorana operators.   It follows that
\begin{equation}
    R_x: \Gamma_{k_n} \rightarrow -(-1)^{n}(-1)^{L+1} \Gamma_{k_n} G,
    \label{rx}
\end{equation}
from which we can infer the reflection properties of many-body states featuring arbitrary numbers of fermion excitations, at least relative to the ground state.  

Crucially, which fermion fillings define physical states descends from boundary conditions of the CFT---not to be confused with the non-negotiable fermion boundary conditions in Eq.~\eqref{BCs}. In our chains the Ising CFT exhibits a unique stable (fixed) boundary condition at which the CDW order parameter is pinned at each edge.  Moreover, the relative sign of the order parameter on the two edges depends on whether the number of sites $L$ is even or odd.  This even-odd effect also originates from the slightly different reflection symmetry preserved by chains with even $L$ (bond centered) versus odd $L$ (site centered).  
Consequently, symmetry dictates that the edge CDW order parameter obeys $\langle \hat{\sigma}_{i+1/2}\rangle = \langle \hat{\sigma}_{L+1/2-i}\rangle\ (i\in\{1, \cdots, L\})$ for odd $L$ but $\langle \hat{\sigma}_{i+1/2}\rangle = -\langle \hat{\sigma}_{L+1/2-i}\rangle$ for even $L$.  

For chains with odd $L$, equality of the non-zero order parameter expectation values on the two ends implies that the bulk of the chain can only support an even number of domain walls.  Recalling that each fermion creates one domain wall, the physical states in this case therefore host an \emph{even} number of fermion excitations.  Consider, as a concrete example, starting from the ground state and then acting with a single fermion operator.  The disorder-parameter string operator carried by that fermion would flip the sign of CDW order parameter at one end, in turn yielding a configuration incompatible with the boundary conditions imposed on the low-energy spectrum; hence such states are excluded.  For even $L$, opposite-sign order-parameter expectation values at the two edges necessitate an odd number of domain walls.  Here physical states accordingly host an \emph{odd} number of fermion excitations.  With fine-tuning one can in principle locate an unstable (free) Ising boundary condition at which translation symmetry breaking at the edges does not generate appreciable CDW order.  Domain-wall numbers are then unconstrained at low energies, and the physical spectrum contains both even \emph{and} odd numbers of fermion excitations.

Applying CFT rules provides an alternative path to derive the energy spectrum of the finite-size Ising critical Rydberg chain. The Ising CFT is characterized by a central charge $c=1/2$ and three primary fields $\mathbb{I}_A,\sigma_A,\varepsilon_A$  ($A=R,L$ denotes their chirality) with respective chiral scaling dimensions of $0,1/16$, and $1/2$.  In this language the chiral fermion fields $\gamma_{R,L}$ defined earlier correspond to $\varepsilon_R$ and $\varepsilon_L$ (spin-1/2 and dimension $1/2$). One can also combine the chiral primaries to obtain the CDW order parameter field  $\sigma\sim\sigma_R\sigma_L$ (dimension $1/8$) and a symmetric field $\varepsilon\sim\varepsilon_R\varepsilon_L$ (dimension 1) that moves the chain off of the second-order Ising phase boundary.  

Each primary operator generates a conformal tower with energy levels
\begin{align}\label{eq:spectrum}
E_{\alpha,J} \sim \frac{\pi \hbar v}{L}\left({h_\alpha} + J-\frac{c}{24}\right),~~~~~~J \in \mathbb{Z}_{\geq0},
\end{align}where $h_\alpha$ denotes the chiral dimension of the primary labeled by $\alpha$. Hereafter, we neglect the central charge contribution, which provides an overall energy shift that our experiment does not resolve. 
Boundary conditions determine the allowed primaries and thus the operators that manifest in the spectrum. The operator content of a given boundary fixed point can be derived from the fusion rules of the primary fields corresponding to `boundary states'.  Consider first fixed boundary conditions, and let $\ket{+}\equiv \ket{\tilde{\mathbb{I}}}$ and $\ket{-}\equiv \ket{\tilde{\varepsilon}}$ represent two Cardy states corresponding to the primary fields $\mathbb{I}$ and $\varepsilon$, respectively.  Physically, $\ket{+}$ and $\ket{-}$ represent CDW order parameter pinnings with $+$ and $-$ signs at a particular edge.  The allowed primaries $\alpha$ in Eq.~\eqref{eq:spectrum} follow from the possible outcomes of fusing the fields associated with the Cardy states for each end of the chain.  
When the CDW order parameter takes the same sign at each edge (representing boundary conditions typically labeled $(+,+)$ and $(-,-)$), the fusion rules $\mathbb{I}\times \mathbb{I}=\mathbb{I}$ and $\mathbb{\varepsilon}\times \mathbb{\varepsilon}=\mathbb{I}$ indicate that only $\alpha = \mathbb{I}$ with $h_\mathbb{I} = 0$ appears.  The spectrum correspondingly reads $E_{\mathbb{I},J}\sim \frac{\pi \hbar v}{L} J$.  Modulo the energy offset from the central charge, this spectrum \emph{almost} agrees with the low-lying levels specified for odd $L$ in Table~\ref{tab:parity-side-by-side}, i.e., the $J = 1$ level is missing from the latter.  The $J = 1$ level indeed disappears because the corresponding state created by acting with the generators of the conformal transformations---Virasoro generators---has zero norm.  (The irreducible representations of the Virasoro algebra are obtained by identifying only the states with non-zero norm.)
When the order parameter carries an opposite sign on the two edges (boundary conditions $(+,-)$ and $(-,+)$), the fusion rule $\mathbb{I}\times \varepsilon=\varepsilon$ dictates that only $\alpha = \varepsilon$ with $h_{\varepsilon} = 1/2$ appears.   The energy spectrum follows as $E_{\varepsilon,J}\sim \frac{\pi \hbar v}{L} (1/2+J)$ in harmony with the even-$L$, odd-fermion-number spectrum from Table~\ref{tab:parity-side-by-side}.
Finally, for unstable free boundary conditions, the relevant Cardy state is $\ket{0}\equiv \ket{\tilde{\sigma}}$.  The fusion rule $\sigma\times \sigma=\mathbb{I}+\varepsilon$ implies that $\alpha = \mathbb{I}$ \emph{and} $\alpha = \varepsilon$ appear; the spectrum predicted by Eq.~\eqref{eq:spectrum} then consists of the two conformal towers $E_{\mathbb{I},J}$ and $E_{\varepsilon,J}$. 

The CFT framework also allows us to predict the spatial parity of the excited (descendant) states. Each descendant at level $J$ is obtained by acting with the Virasoro raising operators ($L_{-n}$) in all possible combinations such that the total level satisfies $\sum_i n_i = J$. Using the operator–state correspondence, the action of these operators corresponds to taking $J$ spatial derivatives of the primary field, with each additional derivative changing the parity under reflection.
Using this rule, incorporating the parity properties of the primary itself, and remembering the different meaning of reflection for even- and odd-$L$ chains allows one to deduce reflection eigenvalues for energy eigenstates.  This logic reproduces the reflection properties reported in Table~\ref{tab:parity-side-by-side}. 

Whereas $\mathcal{H}_{\rm Ising} $ admits a free-fermion description, the TCI point is governed by a strongly interacting CFT.  We can, nevertheless, again employ Eq.~\eqref{eq:spectrum} and CFT rules to characterize the low-energy TCI levels.  The TCI CFT exhibits central charge $c=7/10$ and hosts six primary fields of chiral dimensions $0, 3/80,1/10, 7/16, 3/5$ and $3/2$ that we will denote by 
$\{\mathbb{I}_A, \sigma_A, \varepsilon_A, \sigma'_A, \varepsilon'_A, \varepsilon''_A\}$ .  Combining left and right movers leads to various fields of interest \cite{slagle_microscopic_2021}.  The leading CDW order parameter field is $\sigma \sim \sigma_R \sigma_L$ now with dimension $3/40$.  The $\sigma'\sim \sigma_R' \sigma_L'$ field shares the same symmetries but is less relevant, with dimension $7/8$.  Fully symmetric fields $\varepsilon \sim  \varepsilon_R \varepsilon_L$ (dimension $1/5$) and $\varepsilon' \sim  \varepsilon_R' \varepsilon_L'$ (dimension $6/5$) correspond to the two allowed relevant perturbations that can drive the system away from the TCI point in Fig.~\ref{fig:isingspectroscopy}a.  Chiral fermion fields are given by $\varepsilon_R' \varepsilon_L$ and $\varepsilon_L' \varepsilon_R$ (spin-1/2, dimension $7/10$) as well as $\varepsilon''_{R/L}$ (spin-$3/2$, dimension $3/2$).

The TCI CFT in our setup hosts free and fixed boundary conditions that both realize stable fixed points. The fixed-boundary-condition Cardy states $\ket{+}\equiv \ket{\tilde{\mathbb{I}}}$ and $\ket{-}\equiv \ket{\tilde{\varepsilon''}}$ here correspond to the primary fields $\mathbb{I}$ and $\varepsilon''$, respectively; as for the Ising CFT, they correspond to $+$ and $-$ CDW order parameter pinnings. For $(+,+)$ or $(-,-)$ boundary conditions relevant for odd-$L$ chains, the fusion rules $\mathbb{I}\times \mathbb{I}=\mathbb{I}$ and $\mathbb{\varepsilon''}\times \mathbb{\varepsilon''}=\mathbb{I}$ imply that only $\alpha = \mathbb{I}$ appears, and hence $E_{\mathbb{I},J}\sim \frac{\pi \hbar v}{L}J$.  The $J = 1$ level does not appear for the same reason explained for the Ising case.  
For $(+,-)$ and $(-,+)$ boundary conditions relevant for even $L$, the spectrum instead follows from the fusion rule $\mathbb{I}\times \varepsilon''=\varepsilon''$; only $\alpha = \varepsilon''$ then appears, so using $h_{\varepsilon''} = 3/2$ the energy spectrum reads $E_{\varepsilon'',J}\sim \frac{\pi v}{L} (3/2+J)$.  Since $\varepsilon''$ corresponds to a fermionic operator, this sector is compatible with the constraint noted earlier that even-$L$ physical states host an odd number of fermionic excitation with opposite-sign fixed boundary conditions.
For the stable free boundary conditions, the Cardy state $\ket{0}\equiv \ket{\tilde{\sigma}'}$ together with the fusion rule $\sigma'\times \sigma'=\mathbb{I}+\varepsilon''$ implies that $\alpha = \mathbb{I}$ and $\alpha = \varepsilon''$ contribute to the spectrum---yielding the two conformal towers 
$E_{\mathbb{I},J}$ and $E_{\varepsilon'',J}$. 

Finally, we consider the unstable fixed point intervening between the stable fixed and free boundary conditions.  In this case the Cardy states are 
$\ket{0+}\equiv\ket{\varepsilon}$ and $\ket{0-}\equiv\ket{\varepsilon'}$.  Yet again, $+$ and $-$ correspond to the sign of the CDW order parameter at a particular edge, while the appended `0' indicates that the edge can also accomodate configurations with vanishing CDW order parameter.  As an example, the intermediate fixed point admits edge configurations $10\ldots$ (`large' edge CDW order parameter) and $00\ldots$ (zero edge CDW order parameter) with appreciable weights.  The number of domain walls---and hence the number of fermionic excitations---is then no longer constrained contrary to the situation with fixed boundary conditions.  For $(0+,0+)$ and $(0-,0-)$ appropriate for odd $L$, the fusion rules $\varepsilon\times \varepsilon= \varepsilon'\times \varepsilon'= \mathbb{I}+\varepsilon'$ yield a spectrum consisting of two conformal towers, 
$E_{\mathbb{I},J}$ and $E_{\varepsilon',J}=\frac{\pi \hbar v}{L}(3/5+J)$.  For $(0+,0-)$ and $(0-,0+)$ appropriate for even $L$, the fusion rule $\varepsilon \times \varepsilon' = \varepsilon + \varepsilon''$ yields two different conformal towers, $E_{\varepsilon,J}=\frac{\pi \hbar v}{L}(1/10+J)$ and $E_{\varepsilon'',J}$.  

Table~\ref{tab:TCI_table} summarizes the low-energy level structure and reflection eigenvalues (obtained as for the Ising case) for the TCI CFT.  

\vspace{3mm}
\noindent\textbf{Dependence of spectroscopy signal on modulation wavevector} 

In the main text we discussed spectroscopy with homogeneous global modulation and with a spatially varying modulation pattern at wavevector $k = \pi/(L-1)$, respectively capturing even- and odd-reflection-parity Ising CFT levels (Fig.~\ref{fig:isingLocal}).  
More broadly, one can ask which energy levels our spectroscopy technique reveals when modulating at a general wavevector $k$.  Since momentum is not a good quantum number for open chains, where translation symmetry is broken, one might naively anticipate that only the reflection properties of the modulation matter qualitatively, rather than the precise spatial profile.  
In this section we theoretically analyze  momentum-dependent modulation spectroscopy and show that, on the contrary, it directly reveals a linear dispersion relation via characteristic $k$ dependence of the signal.

Consider the $k$-dependent modulation Hamiltonian
\begin{equation}
\begin{split}
    \delta \hat{H}_k(t) &= A(t)\hat{K}_k,  \\
    \hat{K}_k &\equiv \sum_{j=-\lfloor \frac{L-1}{2} \rfloor}^{\lfloor \frac{L-1}{2} \rfloor} \cos(kj+\alpha)\hat{n}_j.  
\end{split}
\end{equation} 
where $A(t)$ is a temporal sinusoidal modulation profile (for example, Eq.~\eqref{eq:modulationAmplitude}). Notice that the modulation pattern is even under reflection when $\alpha = 0, \pi$ but odd under reflection when $\alpha = \pm\pi/2$. Ultimately, we are interested in understanding the response,  $\delta\langle \hat{K}_k \rangle$, after applying the $\delta \hat H_k(t)$ modulation, so we focus on the transition matrix elements 
\begin{equation}\label{eq:Kge}
    |K_{ge}|^2 = |\bra{g}\hat{K}_k\ket{e}|^2
\end{equation}
that govern the modulation spectroscopy signal strength; see also Eq.~\eqref{eq:peakHeight}. 

First we evaluate Eq.~\eqref{eq:Kge} analytically at small $k$, where we can expand $\hat{K}_k$ in terms of slowly varying fermion fields and, using the preceding Methods section,  express $\ket{e}$ in terms of fermion operators acting on the ground state $\ket{g}$.  To accomplish the former we expand $\hat{n}_j$ in terms of Ising CFT fields via\cite{slagle_microscopic_2021}
\begin{equation}
   \hat{n}_j \sim (-1)^j c_\sigma \sigma + c_\varepsilon \varepsilon + \cdots;
   \label{eq:nexpansion}
\end{equation}
$c_\sigma, c_\varepsilon$ are non-universal constants while the ellipsis denotes subleading contributions that we henceforth drop.  Neglecting the fast-oscillating $(-1)^j$ term, using $\varepsilon = i \gamma_R \gamma_L$, and taking the continuum limit yields
\begin{equation}
    \hat{K}_k 
    \sim i\int_{-L/2}^{L/2} \mathrm{d}x\, \cos(kx+\alpha) \gamma_R(x)\gamma_L(x).
    \label{eq:Ok}
\end{equation}
Since excited states are naturally expressed in terms of the $\Gamma_{k_n}$ operators defined in Eq.~\eqref{eq:GammaOps}, we will further rewrite $\hat{K}_k$ using the transformation
\begin{align}
    \gamma_R(x) &=  \dfrac{1}{\sqrt{2L}}\sum_{k_n>0} (e^{i k_n x}\Gamma_{k_n} + e^{-ik_n x}\Gamma_{k_n}^\dagger), \\
    \gamma_L(x) &=\dfrac{i}{\sqrt{2L}}\sum_{k_n>0} (-1)^{n+1}(e^{-i k_n x}\Gamma_{k_n} - e^{ik_n x}\Gamma_{k_n}^\dagger).
\end{align}
We thereby obtain 
\begin{widetext}
\begin{equation}\label{gammaProd}
    \hat{K}_k 
    \sim \dfrac{1}{L} \int_{-L/2}^{L/2} \mathrm{d}x\, \cos(kx+\alpha)\sum_{k_m, k_n}(-1)^{n+1}\left[e^{i(k_m-k_n)x}\Gamma_{k_m}\Gamma_{k_n} - e^{i(k_m+k_n)x}\Gamma_{k_m}\Gamma_{k_n}^\dagger + h.c. \right]. 
\end{equation}
\end{widetext}

Focusing for now on odd-length chains, $\ket{g}$ is the vacuum with no finite-energy $\Gamma_{k_n}$ fermions populated; excited states $\ket{e}$ arise from adding even numbers of fermion excitations (see Table~\ref{tab:parity-side-by-side}).  Since Eq.~\eqref{gammaProd} is quadratic in fermion operators, only excited states with exactly two populated fermions can possibly contribute nontrivially to Eq.~\eqref{eq:Kge}.  We therefore specialize to $\ket{e} = \Gamma_{k_a}^\dagger \Gamma_{k_b}^\dagger \ket{g}$ with occupied levels labeled by momenta $k_a$ and $k_b$.  The corresponding transition matrix elements evaluate to
\begin{widetext}
    \begin{equation}
        \begin{split}
            \bra{g}\hat{K}_k \ket{e} &= \bra{g}\hat{K}_k \Gamma_{k_a}^\dagger\Gamma_{k_b}^\dagger\ket{g}\\
            &\sim\dfrac{1}{L}\int_{-L/2}^{L/2} \mathrm{d}x\, \cos(kx+\alpha) \left[(-1)^a e^{i(k_b - k_a)x} - (-1)^b e^{i(k_a - k_b)x}  \right] \\
            &= \dfrac{1}{2} \left\{[(-1)^a e^{i\alpha}-(-1)^b e^{-i\alpha}] \mathrm{sinc}\left[(k-(k_a-k_b))\frac{L}{2}\right] -(a\leftrightarrow b)\right\}.  
        \end{split}
    \end{equation}
\end{widetext}
For $\alpha = 0$ or $\pi$ (reflection-symmetric modulation profile), the quantity $(-1)^a e^{i\alpha}-(-1)^b e^{-i\alpha}$ is only nonzero when $a-b \equiv 1\, (\mathrm{mod}\, 2)$. For $\alpha = \pm \pi/2$ (reflection-antisymmetric profile), instead $(-1)^a e^{i\alpha}-(-1)^b e^{-i\alpha}$ is only nonzero when $a-b \equiv 0\, (\mathrm{mod}\, 2)$.  Any of these nontrivial cases yield
\begin{equation}\label{eq:Oge}
|\bra{g}\hat{K}_k\ket{e}|^2 \sim \left\{\mathrm{sinc}\left[(k-(k_a-k_b))\frac{L}{2}\right]+P(a\leftrightarrow b)\right\}^2
\end{equation}
with $P = +1$ for the reflection symmetric case and $P = -1$ for the reflection antisymmetric case. 

Equation~\eqref{eq:Oge} generally predicts maximal transition amplitudes to excited states with energy $E = \hbar v(k_a + k_b)$ when the modulation wavevector satisfies $|k| = |k_a-k_b| = |a-b|\pi/L$.
However, a special case arises for uniform, reflection-symmetric modulations with $k = 0$: there the transition amplitudes peak for $|k_a-k_b| = \pi/L$.
Even-length chains are amenable to a very similar analysis.  One simply needs to recall that the ground state then has the $k_0$ level occupied, and excitations arise from acting an even number of fermion operators (e.g., $\Gamma_{k_a}^\dagger \Gamma_{k_b}^\dagger$ or $\Gamma_{k_0} \Gamma_{k_b}^\dagger$) on that one-fermion state.  

Now we can explain the energy levels experimentally observed in our reflection-symmetric $k = 0$ global modulation and our reflection-antisymmetric $k=\pi/18$ modulation in Fig.~\ref{fig:isingLocal}c. Under the $k = 0$ global modulation, the ground state predominantly couples to excited states with $|k_a-k_b|=\pi/L$.
More explicitly, these states have $k_b = (b+1/2) \pi/L$, $k_a = (b+3/2)\pi/L$, and $E = 2(b+1)\pi \hbar v/L$, where $b\in \mathbb{N}$. These energies form a ladder with ratios $2:4:6:8:\cdots$ and (within our level of approximation) exhibit identical transition strengths $|K_{ge}|^2$ (Eq.~\eqref{eq:Oge}). In our $k = \pi/18$ modulation we chose $\alpha = \pi/2$ to target only reflection-antisymmetric excited states.  
In this case, the ground state predominantly couples to excited states with $|k_a-k_b|=2\pi/L$, i.e., $k_b = (b+1/2) \pi/L$, $k_a = (b+5/2)\pi/L$, and $E = (2b+3)\pi \hbar v/L$, where again $b\in \mathbb{N}$. The resulting ladder ratios read $3:5:7:\cdots$; all transition matrix elements are identical here too. The uniform spacing of excited state energies, together with identical transition matrix elements, leads to a constant response when we modulate the system with different frequencies. This feature relates to universal scaling of the dynamical structure factor, which we will discuss later.

\begin{figure*}[]
    \centering
    \includegraphics[width=1\linewidth]{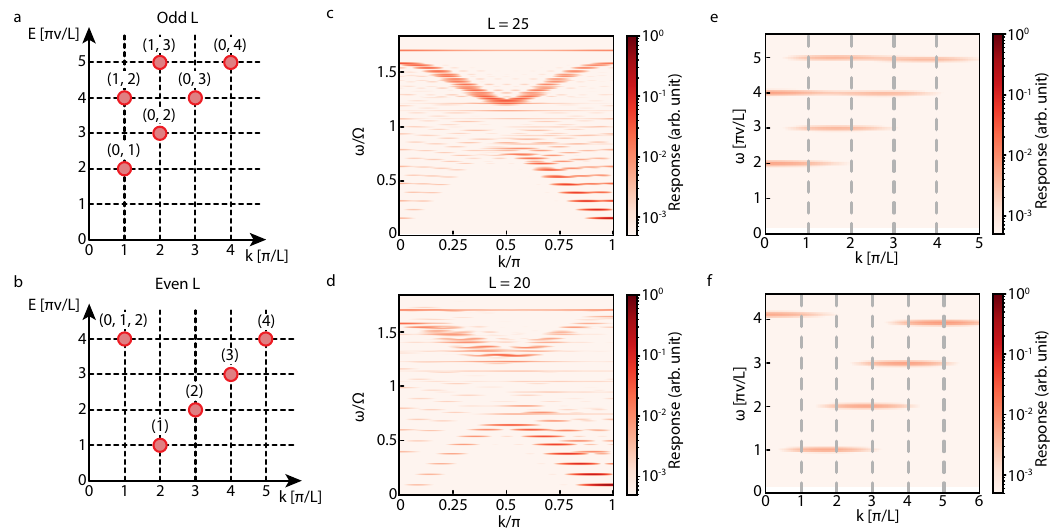}
    \caption{\textbf{Dispersion relation of excited states of a finite-length array.}
    \textbf{(a)} Excited states energies $E$ and their generating momenta $(k_a, k_b)$ for an odd-$L$ array. The $x$-axis $k=|k_a-k_b|$ represents the expected peak positions.
    \textbf{(b)} Excited states energies $E$ and momenta $k$ for an even-$L$ array. 
    \textbf{(c)-(d)} The calculated $k$-dependent response (Eq.~\eqref{eq:kDepResponse}) for (c) an $L=25$ and (d) an $L=20$ array at the critical point with $V_2/\Omega = 0.51$. We observe a linear dispersion close to $k=0$ and $k=\pi$. At $k=0$, the response is approximately constant for low-energy excited states.
    \textbf{(e)-(f)} The low-energy, small-momentum region of the response for (e) $L=25$ and (f) $L=20$. The $x$-axis is expressed in unit of $\pi/L$ and the $y$-axis is expressed in unit of $\pi v/L$, where we choose $v$ to match the first excited state energy. We observe the peak responses appear at positions where we expect excited states (from (a) and (b)). 
    }
    \label{fig:EnergyDispersion}
\end{figure*}

While we did not study other wavevectors experimentally, it is interesting to theoretically explore larger $k$'s.  Equation~\eqref{eq:Oge} predicts that modulation spectroscopy should detect a ``light-cone'' dispersion as $k$ increases: 
When we modulate at $k = m\pi/L$ ($m\in \mathbb{N}^*$), we expect strong response from states with $|k_a-k_b| = m\pi/L$.  The lowest-energy state that one can efficiently probe then has energy $E = \hbar v (m + (-1)^{L+1}) \pi/L = \hbar v k + O(L^{-1})$. Ext. Data Fig.~\ref{fig:EnergyDispersion}a,b shows the analytically predicted excited state energies $E$ that our method can resolve with modulation wavevectors $k = m \pi/L$.  
We also calculate the response
\begin{equation}\label{eq:kDepResponse}
     \dfrac{\pi w^2 A^2}{4} \sum_{e}|K_{ge}|^2 e^{-\frac{w^2}{2}(\omega - \Delta E_{eg}/\hbar)^2}
\end{equation}
as a function of $\omega$ and $k$ directly from the microscopic Rydberg Hamiltonian in Eq.~\eqref{eq:masterH} for 25-atom and 20-atom arrays at $V_2/|\Omega|=0.51$ (Ext.~Data Fig.~\ref{fig:EnergyDispersion}c,d). The anticipated light-cone structure near $k = 0$ clearly manifests.  Additionally, zooming into the small-$k$, low energy part of the response, we find peak positions that are consistent with analytic predictions (Ext.~Data Fig.~\ref{fig:EnergyDispersion}e,f).

Our analytical treatment leading to Eq.~\eqref{eq:Oge} assumed small $k$, allowing us to neglect the fast-oscillating $(-1)^j \sigma$ term in Eq.~\eqref{eq:nexpansion}.   
Near $k = \pi$, the situation flips: the $\varepsilon$ contribution in Eq.~\eqref{eq:nexpansion} becomes unimportant and $\hat{K}_k$ in Eq.~\eqref{eq:Ok} instead involves the lower-scaling-dimension $\sigma$ field.  
Our numerical calculation of the response to modulation (Ext. Data Fig.~\ref{fig:EnergyDispersion}c,d) nevertheless shows that the resolved energy levels are very similar near $k = 0$ and $k = \pi$.  Interestingly, the latter offers a potential advantage that would be worth exploring experimentally in future work: the signal for the lowest-energy states is significantly stronger near $k = \pi$, but diminishes as the energy increases.

\vspace{3mm}
\noindent\textbf{Relation between the dynamical structure factor and the CFT correlation functions}

The connection to field theory follows from the standard correspondence between the temporal response of a one-dimensional quantum Ising model and correlation functions of the associated two-dimensional classical Ising model~\cite{sachdev_quantum_2011}. The imaginary-time two-point correlation function of some field $\hat \phi$ reads 
\begin{equation}\label{eq:finitebeta}
    C(x, \tau)= \mathcal{Z}^{-1} \mathrm{Tr}[e^{-\beta H}\hat{\phi}(x, \tau) \hat{\phi}(0, 0)],
\end{equation}
where $\tau>0$ and $\hat{\phi}(x, \tau) = e^{H\tau} \hat \phi(x)e^{-H\tau}$.
Fourier transforming $C(x,\tau)$ in both space and imaginary time recovers the momentum- and frequency-dependent correlation function
\begin{equation}
    \chi(k, i\omega_n) = \int\mathrm{d}x\int_{0}^{\beta} \mathrm{d}\tau \, e^{i(\omega_n \tau-kx)} C(x, \tau)
    \label{eq:chiinit}
\end{equation}
with $\omega_n =  2\pi n/\beta\, (n\in \mathbb{Z})$ being Matsubara frequencies whose discretization follows from the Euclidean space-time construction enforcing $C(x,\tau) = C(x,\tau+\beta)$. Taking $\beta\to\infty$ and assuming that $\hat \phi$ maps to some scalar primary field $\phi$ with scaling dimension $\Delta_{\phi}$, in the thermodynamic limit we can use the universal scaling law $C(x,\tau) \sim (x^2+v^2\tau^2)^{-\Delta_\phi}$ 
to obtain
\begin{align}
\chi(k, i \omega_n)\sim \left(\sqrt{v^2k^2+\omega_n^2}\right)^{2\Delta_{\phi}-2}.
\end{align}
(We neglect here boundary effects for an open chain but restore them below.)

To relate the imaginary-time correlator with the real-time properties, which are experimentally accessible, we compute the Fourier transform of the real-time correlator ${\tilde{C}(x,t)}$, yielding the dynamical structure factor
\begin{equation}
    S(k, \omega) = \int \mathrm{d}^dx\int_{0}^\infty \mathrm{d}t\, \tilde{C}(x, t)e^{-ikx+i\omega t}.
\end{equation}
By performing an analytic continuation on $\chi(i\omega_n \rightarrow \omega + i0^+$), the dynamical structure factor is then related to $\chi$ by the fluctuation-dissipation relation 
\begin{equation}
\begin{split}
    S(k, \omega) \propto &(1-e^{-\beta\omega})^{-1}\mathrm{Im}(\chi(k, \omega)) \\ 
    \overset{\beta\rightarrow+\infty}{\propto} &(\omega^2-v^2k^2)^{\Delta_{\phi}-1}\theta(|\omega|-v|k|),
    \label{eq:DSF_scaling}
\end{split}
\end{equation}
where $\theta(x)$ denotes the Heaviside step function. 
For the two nontrivial Ising CFT primary fields, $\varepsilon$ and $\sigma$, their scaling dimensions are $\Delta_\varepsilon = 1$ and $\Delta_\sigma = 1/8 $, respectively. Therefore, we expect
\begin{align}
    S_\varepsilon(k, \omega) &\propto \theta(|\omega|-v|k|),\label{eq:SepsilonIsing}\\
    S_\sigma (k, \omega) &\propto (\omega^2-v^2 k^2)^{-\frac{7}{8}}\theta(|\omega|-v|k|).
\end{align}
Here, we find $S_\varepsilon(k=0, \omega)$ to be a constant, which is consistent with the constant response at $k=0$ shown in Ext. Data Fig.~\ref{fig:EnergyDispersion}c. In addition, $S_\sigma(k=0, \omega)$ decays as $\omega^{-\frac{7}{4}}$. This corresponds to the response at $k=\pi$ in Ext. Data Fig.~\ref{fig:EnergyDispersion}c, where the decaying response towards larger $\omega$ qualitatively reflects this power-law scaling.

In principle, to analytically calculate the universal scaling function of the dynamical structure factor of a chain with open boundaries, one should use results for CFT field correlators in the presence of fixed $++$ boundary conditions, which we will show in detail later. Alternatively, we analytically calculate the dynamical structure factor of $\varepsilon$ field using Eq.~\eqref{eq:dynamicalStructureFactor}. By definition, the system's total response is the product of the response contributions from a single energy level, multiplied by the number of energy levels per unit energy $g(\omega)=L/(2\pi v)$. To calculate the contributions from a single level, we can use Eq.~\eqref{eq:Oge} and notice that in the thermodynamic limit ($L\rightarrow\infty$),
\begin{equation}
\begin{split}
    &\sum_{\{\ket{e}:E_e=\omega\}} |\bra{g} \hat{K}_k \ket{e}|^2\\
    \sim&\sum_{\left\{k_a,k_b:k_a+k_b=\frac{\omega}{v}\right\}} \left\{\mathrm{sinc}\left[(k-(k_a-k_b))\frac{L}{2}\right]+P(a\leftrightarrow b)\right\}^2\\
    =&\left\{\begin{split}
        &2,\, k=0,\\
        &1,\, |k|>0\text{ and } |k|< \frac{\omega}{v},\\
        &0,\, |k|> \frac{\omega}{v},
    \end{split}\right.
\end{split}
\end{equation}
converges to a constant and is independent of $L$. Hence, the zero-temperature $\varepsilon$-field dynamical structure factor is
\begin{equation}\label{eq:Skomega}
    S(k, \omega) = C\cdot \theta(|\omega|-v|k|),
\end{equation}
where $C$ is a constant, independent of $\omega$ or $L$. This verifies the expectation that $S(k,\omega)$ is constant as long as the modulation frequency creates an excitation above the light cone dispersion.

Alternatively we will show from a direct field-theoretic derivation of the $\varepsilon$ dynamical structure factor that in this specific critical system, the boundary does not significantly alter the scaling arising from the bulk contribution to the dynamical structure factor. When the boundary effect is negligible and the bulk contribution dominates, one can experimentally determine the scaling dimension from dynamical structure factor measurements (Eq.~\eqref{eq:DSF_scaling}).

\vspace{3mm}
\noindent\textit{Dynamical structure factor of an open chain}
\vspace{3mm}

Here we calculate the dynamical structure factor of an open chain and show that its leading order behavior is the same as a chain with periodic boundary condition, which is further proven to have the same scaling as the bulk correlator. For an open chain of length $L$, the two-point imaginary time correlator of the $\varepsilon$ field is~\cite{slagle_microscopic_2021}
\begin{widetext}
    \begin{equation}
    C(x_1,x_2,\tau)= \left(\dfrac{\pi}{L} \right)^2 \dfrac{\sin(\frac{\pi}{L}x_1)\sin(\frac{\pi}{L}x_2)}{[\cosh(\frac{\pi v}{L}\tau)-\cos(\frac{\pi}{L}(x_1-x_2))][\cosh(\frac{\pi v}{L}\tau)-\cos(\frac{\pi}{L}(x_1+x_2))]}
\end{equation}
\end{widetext}
where $v$ is the non-universal velocity as appeared in Eq.~\eqref{HIsing}. The above correlator holds at sufficiently long distances where $|x_1-x_2| \gtrsim a$ for some microscopic scale $a$ below which short-distance physics not captured by the CFT kicks in.  As an initial step of calculating the dynamical structure factor at $k=0$, we evaluate the imaginary frequency Fourier transform of $C$:
\begin{widetext}
\begin{equation}
    \chi(i\omega) = \dfrac{2L}{\pi v} \int_0^\infty d\tilde\tau e^{i \tilde \omega \tilde \tau} \int_0^\pi du \int_\lambda^u dy \dfrac{\sin(\frac{u+y}{2})\sin(\frac{u-y}{2})}{(\cosh{\tilde\tau} -\cos{y})(\cosh \tilde\tau-\cos u)}
\end{equation}
\end{widetext}
where $\tilde \tau = \pi v\tau/L$ is the normalized time, $\tilde \omega = \omega/(\pi v/L)$ is the normalized frequency, $u = \pi(x_1+x_2)/L$ represents the distance of the center-of-mass position to the boundary, and $y = \pi(x_1-x_2)/L$ represents the distance between two sites. Here $\lambda = \pi a/L$ is a short-distance cutoff that regulates the divergence of the correlator where the field theory description breaks down. First, we show that the integral over $u$ will not lead to divergence, i.e., the boundary contribution does not alter the scaling of $\chi$. Since the $u$-dependent integral is always bounded by
\begin{equation}
    \dfrac{\sin(\frac{u+y}{2})\sin(\frac{u-y}{2})}{\cosh \tilde\tau-\cos u} < \dfrac{\sin^2(\frac{u}{2})}{1-\cos u} = \dfrac{1}{2},
\end{equation}
the integral can then be rewritten as
\begin{equation}
    \chi(i\omega) = \dfrac{L}{\pi v} \int_0^\infty d\tilde\tau e^{i \tilde \omega \tilde \tau} \int_\lambda^\pi dy \dfrac{1}{\cosh \tilde\tau-\cos y} f(\tilde\tau),
\end{equation}
where $f(\tilde \tau)$ is a continuous and bounded function, given by the following integral:
\begin{widetext}
\begin{equation}
    f(\tilde \tau) = \dfrac{1}{\pi}\int_{0}^\pi d\theta \left[\arctan\left(\dfrac{1-\cosh(\tilde\tau)\cos(\theta)}{\sinh(\tilde\tau)\sin(\theta)}\right)-\arctan\left(\dfrac{-1-\cosh(\tilde\tau)\cos(\theta)}{\sinh(\tilde\tau)\sin(\theta)}\right)\right].
\end{equation}
\end{widetext}
Despite that this function could be expressed using series expansion, we notice that this function roughly scales as $f(\tilde \tau)\approx \pi e^{-\tilde \tau}$. Since the scaling of $\chi$ is dominated by the divergence behavior of the integrand, $f(\tilde \tau)$ as a bounded function will not alter the frequency scaling significantly. Therefore, we ignore its contribution from now on. More systematically, since the integrand is dominated by $\tilde \tau$ near zero, one can approximate $f(\tilde \tau) \approx f(0) = \pi$.  With this approximation, we notice that the remaining integral is a Fourier transform of the periodic chain correlator
\begin{equation}
    C_\mathrm{PBC}(y, \tilde\tau) = \dfrac{1}{\cosh \tilde\tau-\cos y}.
\end{equation}

We now show that in the physically relevant range of $\omega$, the imaginary frequency dynamic susceptibility scales the same way as the bulk CFT correlator. Since $\pi v/L$ is the characteristic energy scale of the energy gap, in the continuum limit, the modulation frequency $\omega$ should be large comparing with this energy scale. Hence, we assume $\tilde \omega \gg 1$ in the following calculation. Furthermore, we define $ \omega_\mathrm{UV} = v / a$ as the highest energy scale, above which the field theory no longer applies. Therefore, we further assume $\omega \ll \omega_\mathrm{UV}$, or equivalently $\tilde \omega \lambda \ll 1$. In these limits, when $\tilde \tau \gtrsim 1$, the integral over $\tau$ is highly oscillating and can be bounded by a constant. For $\tilde \tau \lesssim 1$, the denominator of the integrand can be Taylor expanded, and the difference from the exact integral is again bounded by a constant. We thereby obtain
\begin{equation}
    \chi(i\omega) \sim \dfrac{2L}{v}\int_0^\infty d\tilde\tau e^{i \tilde \omega \tilde \tau} \int_\lambda^\infty dy \dfrac{1}{y^2+\tilde \tau^2},
\end{equation}
which is the Fourier transform of the bulk correlator
\begin{equation}
    C_\mathrm{bulk}(y, \tilde \tau) = (y^2+\tilde\tau^2)^{-1}.
\end{equation}
This integral evaluates to
\begin{widetext}
\begin{equation}
    \chi(i\omega) = \dfrac{L}{v} \left\{\left[-\pi \log(\tilde \omega \lambda) + O(1)\right] + i \left[\dfrac{\pi^2}{2} + O(\tilde \omega \lambda \log(\tilde \omega \lambda))\right]\right\}.
\end{equation}
\end{widetext}

Finally, we perform analytic continuation to obtain the real-frequency correlation function
\begin{widetext}
\begin{equation}
    \chi(\omega+i\eta) = \dfrac{L}{v}\left\{ \left[-\pi \log(-i(\tilde \omega+i\tilde \eta) \lambda) + O(1)\right] + i \left[\dfrac{\pi^2}{2} + O(\tilde \omega \lambda \log(\tilde \omega \lambda))\right] \right\}.
\end{equation}
\end{widetext}
The dynamical structure factor is the imaginary component of the real-frequency $\chi(\omega)$. Therefore, the leading-order term of $S(k=0, \omega)$ is
\begin{equation}
    S(k=0, \omega) = \dfrac{1}{L}\lim_{\eta\rightarrow 0^+}\mathrm{Im}[\chi(\omega + i\eta)] = \dfrac{\pi^2}{v},
\end{equation}
which is independent of $L$ and frequency-independent.

The above analysis shows that the frequency-scaling of the $\varepsilon$-field correlator in an open-boundary critical chain is the same as for a $(1+1)D$ bulk Ising CFT under frequency coarse-graining. In fact, the dynamical structure factor of an open chain can be analytically calculated if we admit the approximate scaling $f(\tilde \tau) \approx \pi e^{-\tau}$. First we calculate the imaginary-frequency periodic chain correlator via performing the Fourier transform of the periodic boundary correlator
\begin{equation}
\begin{split}
    \chi(i\omega) &= \dfrac{L}{v} \int_0^\infty d\tilde\tau e^{i\tilde \omega \tilde \tau} \int_0^\pi dy \dfrac{1}{\cosh (\tilde \tau + \lambda)-\cos y}\\
    &=  \dfrac{2\pi L}{v}\dfrac{e^{-\lambda} {}_{2}{F}_{1} (1,\frac{1-i\tilde \omega}{2},\frac{3-i\tilde \omega}{2};e^{-2\lambda})}{i+\tilde\omega},
\end{split}
\end{equation}
where $\lambda\rightarrow0^+$ is a cutoff to regulate the divergence and ${}_{2}{F}_{1}$ is a hypergeometric function.
Taking its analytic continuation naturally yields
\begin{equation}
    S_\mathrm{PBC}(k=0, \omega)=\dfrac{\mathrm{Im}[\chi(\omega)]}{L} = \dfrac{2\pi^2}{v}\sum_{m=0}^\infty \delta(\tilde\omega-(2m+1)).
\end{equation}
This is a sum of equally spaced $\delta$-functions with the same amplitude, independent of the system size $L$. If we perform frequency coarse-graining, then this function becomes a constant. 

Now we evaluate the dynamical structure factor for an open chain. Using $f(\tilde \tau)\propto e^{-\tilde \tau}$, the open-boundary dynamic structure factor is related to the periodic-boundary case with 
\begin{equation}
\begin{split}
        S_\mathrm{OBC}(k=0, \omega) &=S_\mathrm{PBC}\left(k=0, \omega-\dfrac{\pi v}{L} \right) \\
        &= \dfrac{2\pi^2}{v} \sum_{m=0}^\infty \delta(\tilde\omega - 2(m+1)),
\end{split}
\end{equation}
which has the same coarse-graining frequency dependence. Interestingly, the center of these $\delta$-functions correspond to the predicted eigenenergies of the boundary CFT. This formula can thus be interpreted as a sum of equal contributions from quantized energy levels with energy ratio $2:4:6:\cdots$.

\vspace{3mm}
\noindent\textbf{Numerical simulation of linear response}

To measure the correlations of CFT fields in the experiment with the modulation-probe sequence, we need to choose the modulation $\hat K$ and the observable $\hat Q$ that are a sum of local field operators. To this end, we first identify the lattice counterparts of the primary field $\varepsilon$~\cite{slagle_microscopic_2021}: $\hat{\varepsilon}_{i+\frac{1}{2}} = \hat n_i + \hat n_{i+1}$ (we will drop the $\frac{1}{2}$ in subscripts hereafter). Hence, the global detuning modulation is a $k=0$ $\varepsilon$-modulation
\begin{equation}
    \hat K = \sum_i \hat n_i = \dfrac{1}{2}\sum_i \hat \varepsilon_i.
\end{equation}
After the global modulation, we measure all $\hat n_i$ to reconstruct the $k=0$ $\varepsilon$-field operator $\hat Q = \hat K$.
Specifically, in the continuum regime, the response of $\hat K$ is proportional to the dynamical structure factor of the $\varepsilon$ field at $k=0$.

We compare our experimental results with tensor-network-based numerical simulation of the modulation dynamics~\cite{hauschild_efficient_2018}. We adopt a matrix product state (MPS) representation of the critical state and find the ground state with the density matrix renormalization group (DMRG) method. For an efficient simulation, here we study the dynamics of the FSS Hamiltonian
\begin{equation}
    \hat{H} = \dfrac{\Omega}{2}\sum_i \hat{P}_{i-1} \hat{X}_i \hat{P}_{i+1} - \Delta \sum_i \hat{n}_i + V_2 \sum_i \hat{n}_i \hat{n}_{i+2}.
\end{equation}
(Truncation to interactions with distance $|i-j| \le 2$ leads to faster numerics.) We use the time-dependent variational principle (TDVP) method to calculate the evolution of the state under a time-dependent Hamiltonian $\hat{H}_0 + A(t) \hat K$. In the end, we calculate the expectation value of the observable $\langle \hat K \rangle.$

Applying the numerical simulation to a 19-atom array, we compare the numerically simulated response with the experiment (Fig.~\ref{fig:dynamicSusceptibility}a). Since the simulated Hamiltonian is different from the system Hamiltonian, we rescale the frequency of the simulated response to match the first spectral peak center with the experiment and find qualitative agreement on the peak heights of the $\sum_i \hat \varepsilon_i$ response.

For the experiment performed in the continuum limit, to show that the nearly constant response at low frequencies in Fig.~\ref{fig:dynamicSusceptibility}b is a feature even in the thermodynamic limit, we numerically simulate the dynamics for system sizes from $L=35$ to $L=85$. We rescale the responses by $(AL)^{-1}$ to retrieve the zero-temperature dynamical structure factor and find that the response functions collapse to a universal function (Fig.~\ref{fig:dynamicSusceptibility}c). The remaining oscillation observed in the response is a result of the finite modulation time and is not a physical feature of the dynamical structure factor.

\vspace{3mm}
\noindent\textbf{Local detuning modulation}

The local detuning control is an essential component in measuring the wavevector-dependent spectrum and  dynamical structure factor, and the $\sigma$-field correlation. In this section, we outline our protocol of synchronizing the trapping tweezer light and the global Rydberg laser detuning modulation to measure the $k$-dependent spectrum. Furthermore, we will explain additional experimental capability needed to measure the $\sigma$-field dynamical structure factor.

To target the $k$-dependent spectrum, we modulate the system with $\delta \Delta_j \propto \cos(kj+\alpha)$. To achieve this, we utilize an AOD to generate a fixed spatial pattern of tweezer intensities and an AOM to control and modulate the overall intensity of the tweezers. The local detuning applied by the individual tweezer beam, which is proportional to the intensity, is 
\begin{equation}
    \delta \Delta_{j\text{, tw.}} = A(t) (c+\cos(kj+\alpha)),
\end{equation}
where $A(t)$ is always negative and $c\ge 1$ due to the sign of the relative light shift between the ground state and the Rydberg state at the trapping wavelength, $813$~nm. We then synchronize the Rydberg beam to apply a global detuning
\begin{equation}
    \delta \Delta_{\text{gl.}} = -c\,A(t).
\end{equation}
Therefore, the overall local detuning is 
\begin{equation}
    \delta \Delta_j = \delta \Delta_{j\text{, tw.}} + \delta \Delta_{\text{gl.}} = A(t)\cos(kj+\alpha).
\end{equation}
When acquiring the data presented in Fig.~\ref{fig:isingLocal}, the global intensity modulation is chosen to be
\begin{equation}
    A(t) = A_0 f(t)(1+\cos(\omega t + \varphi)),
\end{equation}
where $A_0$ is negative and $f(t)$ is the Gaussian envelope function.

To target the $\sigma$-field dynamical structure factor, the required modulation pattern is
\begin{equation}
    \delta \Delta_j = A_0 f(t) \cos(\omega t + \varphi) \cos(\pi j + \alpha),
\end{equation}
where $f(t)$ is an envelope function. The temporal part of this modulation must be chosen to be both positive and negative. Hence, we plan to combine two sets of tweezers and the global detuning modulation
\begin{equation}
\begin{split}
        \delta \Delta_{j\text{, tw. 1}} &= A_0 f(t)(1+\cos(\omega t+\varphi))(c_1+\cos(\pi j+\alpha)),\\
        \delta \Delta_{j\text{, tw. 2}} &= -A_0 f(t)(c_2+\cos(\pi j+\alpha)),\\
        \delta \Delta_{\text{gl.}} &= (c_2-c_1) A_0 f(t) \cos(\omega t+\varphi),
\end{split}
\end{equation}
In the future, we plan to combine two sets of tweezers at 515~nm~\cite{cooper_alkaline-earth_2018} and 813~nm, respectively, to measure the $\sigma$-field dynamical structure factor.

\vspace{3mm}
\noindent\textbf{Quench dynamics at the TCI point}

\begin{figure*}[]
    \centering
    \includegraphics[width=0.5\linewidth]{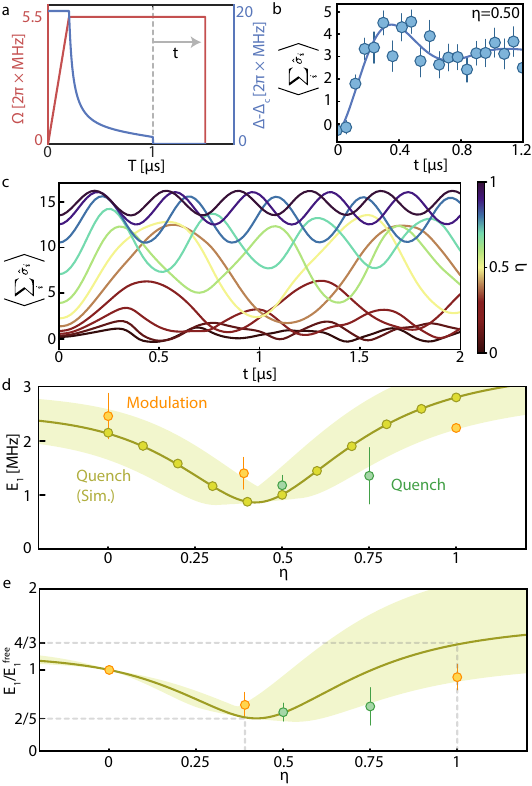}
    \caption{\textbf{Quench dynamics at the tricritical point and the first excited state energy.}
    \textbf{(a)} The experimental sweep. We adiabatically ramp the detuning close to the critical point and then quench it to $\Delta_c$. After waiting for time $t$, we measure the local observables $\hat{n}_i$.
    \textbf{(b)} The oscillation of $\sum_i \hat{\sigma}_i$ after quenching to the tricritical point of a $L=19$ array. The solid line is a fit to $a+b \cos(\omega t+\varphi) e^{-t/\tau}$. The fit frequency $\omega$ is interpreted as the first excited state energy $E_1$.
    \textbf{(c)} Numerical simulation of $\sum_i \hat{\sigma}_i$ after quenching for different detuning strengths $\eta \in[0, 1]$.
    \textbf{(d)} Energies of the first excited state of $L=19$ array under different boundary detunings. Orange and green data points are extracted from modulation and quench data, respectively. Yellow data points are fits of the numerical simulation of the quench dynamics. The solid line is the result of exact diagonalization, with the shaded area indicating the uncertainty due to $V_2$.
    \textbf{(e)} Ratio between the first excited state energy at different $\eta$ and $\eta = 0$ (free boundary condition). When $\eta$ increases from 0, the measured first excited state energy first decrease to $0.57(15)$ at $\eta = 0.39$ and increase to $0.91(15)$ at $\eta = 1$, which agree with the numerical simulation taken uncertainty of $V_2$ into account (shaded area). The solid line is the result of exact diagonalization, where we observe the universal ratio at $\eta = 0.39$ and $\eta = 1$. 
    }
    \label{fig:quench}
\end{figure*}

Besides the equilibrium spectrum, interest in investigating the non-equilibrium dynamics in a quantum system at criticality motivates the observation of quench dynamics~\cite{calabrese_quantum_2016}.
In particular, since TCI CFT spectra have a set of rational fraction, the system will return to its initial state periodically, and physical observables should exhibit oscillations. 
For a low-energy initial state, the dominant frequency component of these oscillations should reflect the excitation gap, which provides a complementary approach to measure the first excited state energy~\cite{manovitz_quantum_2025}.
Here we prepare a non-equilibrium low-energy TCI state by adiabatically ramping our system close to the TCI point and then quenching the detuning to $\Delta_c$ (Ext. Data Fig.~\ref{fig:quench}a). After holding at the TCI point for different times, we measure the sum of local fields, $\sum_i \hat{\sigma}_{i}$, and observe damped oscillations (Ext. Data Fig.~\ref{fig:quench}b). We record the oscillations at boundary detunings $\eta = 0.5, 0.75$ and fit the data to a trial function $a+b \cos(\omega t+\varphi)e^{-t/\tau}$; the fit $\omega$ is interpreted as the excitation energy gap $E_1$. To verify the relation between the oscillation frequency and the energy gap, we numerically simulate the quench dynamics for different local detuning strengths $\eta$. For all $\eta\in [0, 1]$, we observe long-lived oscillations (Ext. Data Fig.~\ref{fig:quench}c), with the fitted frequencies consistent with the first excited state energy $E_1$ (yellow data points in Ext. Data Fig.~\ref{fig:quench}d), indicating an oscillation between the CFT ground state and the first excited state.

We then extract the energy gap from the experimental data and compare with the numerically calculated first excited state energy. We find the fitted oscillation frequencies are consistent with the energy gaps of the Hamiltonian (green data points in Ext. Data Fig.~\ref{fig:quench}d), up to the uncertainty in $V_2$. Despite the fact that we observed a damped oscillation, mainly due to noise-induced decoherence, we suspect that the coherence time can be improved by suppressing laser noises. Together with the modulation spectroscopy (orange data points in Ext. Data Fig.~\ref{fig:quench}d), all measured $E_1$ values are consistent with the numerical calculation, up to experimental uncertainties. This agreement further supports the claim that we tune the boundary conditions by applying local detunings at the TCI point.

\section*{Acknowledgements}

We thank Joonhee Choi, Soonwon Choi, Roland Farrell, Paul Fendley, Ran Finkelstein, Yue Liu, Nandagopal Manoj, Daniel Mark, Pablo Sala, Thomas Schuster, Adam Shaw, Ning Su, Federica Surace, and Cenke Xu for insightful discussions, as well as Elie Bataille, Kon Leung, Hannah Manetsch, and Gyohei Nomura for technical support of the experimental setup. We also thank Erin Burkett and Divesh Soni for feedback to the manuscript. This material is based upon work supported by the U.S. Department of Energy, Office of Science, National Quantum Information Science Research Centers, Quantum Systems Accelerator. Additional support is acknowledged from the Heising-Simons Foundation (2024-4852), AFOSR (FA9550-23-1-0625), NSF QLCI program (2016245), Institute for Quantum Information and Matter, an NSF Physics Frontiers Center (NSF Grant PHY-2317110), DOE (DE-SC0021951), Army Research Office MURI program (W911NF2010136), NSF QLCI program (OMA-2016245), and Technology Innovation Institute (TII). The U.S. Department of Energy, Office of Science, National Quantum Information Science Research Centers, Quantum Science Center supported theoretical analysis of this work.
SM acknowledges the Walter Burke Institute for Theoretical Physics at Caltech. 
YL acknowledges support from AWS Quantum Postdoctoral Fellowship. RBST acknowledges support from the Taiwan-Caltech Fellowship. LP acknowledges support from the David and Ellen Lee Postdoctoral Prize Fellowship. MK acknowledges support from the Deutsche Forschungsgemeinschaft (DFG, German Research Foundation) under Germany’s Excellence Strategy–EXC–2111–390814868, the European Union (grant agreement No 101169765), as well as the Munich Quantum Valley, which is supported by the Bavarian state government with funds from the Hightech Agenda Bayern Plus. 

\begin{table*}[]
  \centering
  \footnotesize
  \setlength{\tabcolsep}{5pt}
  \renewcommand{\arraystretch}{1.15}
  \setlength{\arrayrulewidth}{0.6pt}

  \begin{tabular}{|c|c|c|c|c|c|}
    \hline
    \multicolumn{3}{|c|}{\textbf{Odd $L$}} &
    \multicolumn{3}{c|}{\textbf{Even $L$}} \\
    \cline{1-3}\cline{4-6}
    \textbf{Occupied $n$} & \textbf{$E \times L/(\pi\hbar v)$} & \textbf{$R_x$ parity} &
    \textbf{Occupied $n$} & \textbf{$E \times L/(\pi\hbar v)$} & \textbf{$R_x$ parity} \\
    \hline
    $\varnothing$ (ground state)          & $0$   & even &  $\{0\}$ (ground state)             & $1/2$ & even \\
    $\{0,1\}$              & $2$   & even & $\{1\}$               & $3/2$ & odd  \\
    $\{0,2\}$              & $3$   & odd  & $\{2\}$               & $5/2$ & even \\
    $\{0,3\},\ \{1,2\}$    & $4$   & even & $\{3\}$               & $7/2$ & odd  \\
    $\{0,4\},\ \{1,3\}$    & $5$   & odd  & $\{0,1,2\},\ \{4\} $   & $9/2$ & even \\
    \hline
  \end{tabular}

    \caption{\textbf{Lowest many-body energies $E$ for Ising critical Rydberg chains with odd- and even-$L$.} These eigenstates correspond to even and odd fermion-parity sectors, respectively.   ``Occupied $n$'' lists the set of occupied single-particle fermionic modes. The ground-state reflection parities are verified using exact diagonalization; the excited-state reflection parities are then determined from Eq.~\eqref{rx}.} 
  \label{tab:parity-side-by-side}
\end{table*}

\begin{table*}[]
  \centering
  \footnotesize
  \setlength{\tabcolsep}{4pt}
  \renewcommand{\arraystretch}{1.15}
  \setlength{\arrayrulewidth}{0.6pt}

  \begin{tabular}{|c|c|c|c|c|c|c|c|c|}
    \hline
    \multicolumn{9}{|c|}{\textbf{Odd $L$}} \\ \hline

    \multicolumn{3}{|c|}{\textbf{Free}} &
    \multicolumn{3}{c|}{\textbf{Intermediate}} &
    \multicolumn{3}{c|}{\textbf{Fixed}} \\ \hline

    \textbf{Primary} & \textbf{$E \times L/(\pi\hbar v)$} & \textbf{$R_x$ parity} &
    \textbf{Primary} & \textbf{$E \times L/(\pi\hbar v)$} & \textbf{$R_x$ parity} &
    \textbf{Primary} & \textbf{$E \times L/(\pi\hbar v)$} & \textbf{$R_x$ parity}
    \\ \hline

    $\mathbb{I}$ & $0$   & even & $\mathbb{I}$ & $0$       & even & $\mathbb{I}$ & $0$ & even \\
    $\varepsilon''$ & $3/2$ & even & $\varepsilon'$ & $3/5$     & even & $\mathbb{I}$ & $2$ & even \\
    $\mathbb{I}$ & $2$   & even & $\varepsilon'$ & $1+3/5$ & odd  & $\mathbb{I}$ & $3$ & odd  \\
    $\varepsilon''$ & $5/2$ & odd  & $\mathbb{I}$ & $2$       & even & $\mathbb{I}$ & $4$ & even \\ \hline

  \end{tabular}

  \vspace{1em}

  \begin{tabular}{|c|c|c|c|c|c|c|c|c|}
    \hline
    \multicolumn{9}{|c|}{\textbf{Even $L$}} \\ \hline

    \multicolumn{3}{|c|}{\textbf{Free}} &
    \multicolumn{3}{c|}{\textbf{Intermediate} } &
    \multicolumn{3}{c|}{\textbf{Fixed}} \\ \hline

    \textbf{Primary} & \textbf{$E \times L/(\pi\hbar v)$} & \textbf{$R_x$ parity} &
    \textbf{Primary} & \textbf{$E \times L/(\pi\hbar v)$} & \textbf{$R_x$ parity} &
    \textbf{Primary} & \textbf{$E \times L/(\pi\hbar v)$} & \textbf{$R_x$ parity} \\ \hline

    $\mathbb{I}$ & $0$   & even & $\varepsilon$  & $1/10$   & even & $\varepsilon''$ & $3/2$ & even \\
    $\varepsilon''$ & $3/2$ & odd  & $\varepsilon$  & $1+1/10$ & odd  & $\varepsilon''$ & $5/2$ & odd \\
    $\mathbb{I}$ & $2$   & even & $\varepsilon''$ & $3/2$    & even & $\varepsilon''$ & $7/2$ & even \\
    $\varepsilon''$ & $5/2$ & even & $\varepsilon$  & $2+1/10$ & even & $\varepsilon''$ & $9/2$ & odd \\ \hline
  \end{tabular}

  \caption{\textbf{Lowest many-body energies $E$ and reflection parities $R_x$ for odd- and even-$L$ TCI critical Rydberg chains realizing the free, intermediate, and fixed boundary conditions. }
  }
  \label{tab:TCI_table}
\end{table*}

\end{document}